\documentclass[
aps,%
12pt,%
final,%
notitlepage,%
oneside,%
onecolumn,%
nobibnotes,%
nofootinbib,%
groupedaddress,%
noshowpacs,%
centertags]%
{revtex4}
\usepackage [cp866]{inputenc}
\usepackage [english]{babel}
\usepackage {caption2}
\usepackage{epstopdf}

\def\bge{\begin{equation}}
\def\ene{\end{equation}}
\def\bgea{\begin{eqnarray}}
\def\enea{\end{eqnarray}}

\begin{document}
\begin{center}
{\Large \bf{Triplet production by a polarized photon beam on a polarized
electron target}}

\vspace{4mm}

G.I.~Gakh$^{(a, b)}$, M.I.~Konchatnij$^{(a)}$, N.P.~ Merenkov$^{(a, b)}$

\vspace{4 mm}
%$^{(a)}$Jefferson Lab, Newport News, VA 23606, USA\\
%$^{(b)}$Duke University,
%Durham, NC 27708, USA \\
%$^{(c)}$ Center of Particle and High Energy Physics,
%220040 Minsk, Belarus \\

$^{(a)}$ NSC ''Kharkov Institute of Physics and Technology'',\\
Akademicheskaya, 1, 61108 Kharkov, Ukraine \\
$^{(b)}$ V.N.~Karazin Kharkiv National University, 61022 Kharkov,
Ukraine
\end{center}
\vspace{0.5cm}

\begin{abstract}
Polarized and unpolarized observables in the process of the triplet
photoproduction on a free electron, $\gamma +e\to e+e+e$, have been
calculated in the laboratory system. These observables are
calculated in the approximation when the Borsellino and $\gamma -e$
diagrams are taken into account. Using the method of the invariant
integration over the produced electron-positron pair variables, the
different distributions were obtained in the analytical form. We
obtain the double distribution over the $q^2$ (the square of the
four-momentum transfer to the recoil electron) and $Q^2$ (the
created $e^+e^-$-pair invariant mass squared) variables, and single
distributions over $q^2$ or $Q^2$ variables. We consider the
following cases: unpolarized particles, the photon beam is linearly
polarized and circularly polarized photon beam interacts with a
polarized electron target. The influence of the $\gamma -e$ terms on
the calculated observables has been investigated. The possibility to
use this process for the measurement of the circular polarization of
the high-energy photon beam is also considered. The numerical
estimations of various polarization observables have been done.
\end{abstract}

\maketitle

\section{Introduction}
\hspace{0.7cm}

Many authors have studied theoretically the process of the triplet
photoproduction (TPP) by an unpolarized photon on free electrons.
This process is completely described by 8 Feynman diagrams.
Therefore, the exact expressions for the differential and partly
integrated cross sections of the TPP process are very cumbersome and
exist in the complete form only in the unpolarized case \cite{H75}.
That is why, the usual practice, in the calculations of various
observables, was to use approximations: for example, to consider the
high-energy limit or to neglect some of the Feynman diagrams. The
approximation, when only the so-called Borsellino diagrams (Fig. 1a)
are taken into account, is the most often used in the calculations.

Borsellino \cite{B47} calculated the cross section of the TPP
process by neglecting the exchange and $\gamma - e$ diagrams. He
derived formula for the total cross section in the high-energy
limit. Mork \cite{M67} calculated numerically the total cross
section of the TPP reaction. He calculated separately the
contributions to the total cross section from the full set of the
Feynman diagrams (Borsellino, $\gamma - e$ and exchange terms). It
was found that the Borsellino's calculation \cite{B47} for the total
cross section is valid for $E_{\gamma}\ge $8 MeV. The calculated
recoil-electron momentum distribution is also agrees with the
Borsellino result which is valid if the exchange and $\gamma - e$ terms
can be neglected. Mork showed that the Borsellino's distribution is
valid for $E_{\gamma}\ge $8 MeV if the energy of the recoil electron
is well below the energy of the electron produced. Because Mork took
into account all diagrams, he has verified the accuracy and
limitations of the previous calculations. Haug \cite{H75} took into
account all diagrams and calculated analytically the angular
distribution and energy spectra of the produced positron as well as
the total cross section. Later \cite{H81}, the total cross section
is fitted to simple analytic expressions in four intervals of the
incident photon energy which cover the entire range between threshold
and infinity. Endo and Kobayashi \cite{EK93} calculated numerically
the differential cross section of the TPP on an electron target in
the photon energy region $E_{\gamma}=50 - 550$ MeV. They took into
account all eight Feynman diagrams, i.e., the calculation was done
without any approximation. For the recoil-electron momentum
distribution, it turned out that the calculation with only the
Borsellino diagrams agrees very well with the full calculation. The
total cross sections for the process of $n$ electron-positron pair
production in photon-electron collisions have been calculated at
high energy in the main logarithmic approximation \cite{AEA08}. In
the work \cite{ID14}, the authors have analyzed 8 Feynman diagrams
and have shown that for energies lower to $\sim $ 500 MeV, the
assumption about clear distinction between recoil electron and pair
electron is not a good approximation. They proposed the way to solve
this problem.

The authors of Ref.~\cite{BFK66} obtained the expression for the
Compton tensor of the fourth rank, which determines the Borsellino
diagrams, in the case when the integration over the produced
electron-positron pair variables has been performed using the method
of the invariant integration. This calculation was done for the
general case of two virtual photons.

A few papers were devoted to the investigation of the polarization
effects in the triplet photoproduction on an electron. The authors
of the paper \cite{BP71} calculated the differential cross section
in the azimuthal angle of the recoil electrons in the case of the
linearly polarized photons taking into account the Borsellino
diagrams only. The cross section asymmetry slowly decreases from 30
$\%$ at the 10 MeV of the photon beam energy down to 14 $\%$ in the
far asymptotic. Later, they \cite{BP74} investigated the influence
on the asymmetry of the minimal detected recoil momentum. The
possibility of the determination of the degree of a linearly
polarized photon beam with the help of the triplet photoproduction
was studied also in Refs. \cite{VK72, VM75}. The differential cross
section with respect to the azimuthal angle is determined
\cite{VK72} for extremely high photon energies in Lab. system. The
distribution with respect to the polar angle of the recoil electrons
is discussed also in this paper. The authors of Ref.~\cite{VM75}
calculated the recoil electron distribution and the energy
distribution of one of the pair components as well as the positron
(electron) spectrum in the case when the recoil momentum is being
fixed. The authors of  Ref. \cite{EK93} calculated the variation of
the cross section with the azimuthal angle of the recoil electron
and obtained the analyzing power for the polarimetry of linearly
polarized photons. It was found that the analyzing power is greatly
enhanced by selecting the events with small opening angles between
the forward going electron and positron. The detailed description of
the different differential distributions, such as the dependence on
the momentum value, on the polar angle and minimal recorded momentum
of the recoil electron, dependence on the invariant mass of the
created electron-positron pair, on the positron energy and others,
has been investigated in Ref.~\cite{BVMP94}. The authors of
Ref.~\cite{AEA00} calculated the power corrections of the order of
$m/\omega $ (m is the electron mass and $\omega $ is the energy of
the photon beam in the laboratory system), but only due to the
interferences of the Borsellino diagrams with all the rest ones, to
the distribution of the recoil-electron momentum and azimuthal
angle. They estimated the deviation from the asymptotic result
\cite{VK72} for various values of $\omega $. The possibility of the
determining the circular polarization of a high-energy photon beam
by the measuring the created electron polarization was investigated
in Ref.~\cite{GKLM13}, taking into account the contribution of the
Borsellino diagrams. Different double and single distributions of
the created electron were calculated.

Note that the polarization effects in the TPP process were
calculated, up to now, in the approximation when only the
Borsellinos' diagrams were taken into account. Such approximation
was checked by many authors for the calculated unpolarized
observables such as the total cross section, the distribution over
the recoil-electron variables. It was found that such approximation
is valid for the photon-beam energies $\geq $ 10 MeV. But such
verification was not performed for the investigated polarization
observables in the TPP process. That is why, in this paper, we
calculated the polarization observables taking into account not only
the Borsellinos' diagrams but we take also into account the $\gamma
-e$ diagrams.

The results of the paper \cite{BP71} stimulated a search for the
construction of a good polarimetry for high-energy photons above an
energy of a few hundred MeV. So, for the measurement of the
photon-beam linear polarization it was worthwhile to develop
polarimetry on the basis of the detection of the recoil electrons
from the TPP process. For example, the authors of Ref.~\cite{EAT89}
constructed a scintillation counting system for detecting the
recoils electrons in TPP process. It was tested by using tagged
photons with $E_{\gamma}=120 \div 400$ MeV and probed to be capable
of identifying the TPP events with recoil-electron momenta 1.92 - 10
MeV/c.

The astrophysics community shows interest in measuring the
polarization of the cosmic gamma rays of the high energy: for
example, the proposal Hard X and Gamma-ray Polarization (see
references in \cite{DI09}). The problem of the measurement of the
$\gamma\gamma $ and $\gamma e$ luminosities and polarizations at
photon colliders has been considered in Ref.~\cite{PEA04}. There are
two QED processes are of interest to measure the $\gamma e$
luminosity: the Compton scattering $\gamma +e\to\gamma +e$ and the
TPP reaction on free electron $\gamma +e\to e+e+e$. At small angles
the TPP cross section is even larger and this cross section only
weakly depends on the polarization of the initial particles. The
study of the TPP process on a free electron is a part of the
proposed physics program at the planned high brightness linac IRIDE
(Interdisciplinary Research Infrastructure based on Dual Electron
linac and laser), Frascati  \cite{IRIDE}. At IRIDE one can search
for a new, beyond Standard Model, weakly interacting U boson (see,
for example, \cite{FAP}). Its existence can explain several puzzling
astrophysical observations (PAMELA abundance of positrons and
others). The mass of the U boson is expected to be at MeV or GeV
scale. Thus, the QED triplet production is the main background in
the search of U boson in the electron-photon scattering that
requires the precise measurement of the triplet production. This, in
its turn, needs the precise theoretical calculations for the various
observables in the TPP process.

In this paper we calculate polarized and unpolarized observables in
the TPP process on a free electron in the approximation when the
Borsellino and $\gamma -e$ diagrams are taken into account.
Integrating the differential cross section over the produced
electron-positron pair variables, using the method of the invariant
integration, we obtain the analytical form of the distribution over
the recoil-electron variables. These are the squared invariant mass
of the created $e^+e^-$ pair $(p_1+p_3)^2$, and squared momentum
transferred $(p_2-p)^2$. For the definition of the particle
4-momenta see Fig. 1. Really, our results describe the events with
well separated created and recoil electrons. Otherwise, the effects
due to the identity of the final electrons have to be taken into
account. In this case, the investigation of double distributions
over both $(p_1+p_3)^2$ and $(p_2+p_3)^2$ variables seems more
natural. We believe that search for U-boson at IRIDE can be
performed in the kinematical regions where the contribution of the
$\gamma -e$ diagrams is at least of the same order as the Borsellino
ones. Besides, we understand that at this condition the identity
effects have become essential. Thus, for the U-boson search our
results have to be recalculated. Nevertheless, they can be used to
determine the optimal kinematical region and to test the Monte Carlo
generators which used for analysis of the real experiments. In what
follows we consider the following cases: unpolarized particles, the
photon beam is linear polarized and circularly polarized photon beam
interacts with a polarized electron target. The influence of the
$\gamma -e$ terms on the azimuthal asymmetry, due to the linearly
polarized beam, and on the double-spin asymmetry, caused by the
circularly polarized photon beam and electron-target polarization,
has been investigated. The analytical expressions are obtained for
various observables and the numerical estimations of the
polarization effects have been done. The most of our analytical
results are absent in the literature.

In Section 2 the matrix element describing the Borsellino and
$\gamma -e$ diagrams is given. In Section 3 the tensors defining the
Borsellino diagrams were calculated and the distributions over the
$q^2, Q^2$ and $\varphi $ (the azimuthal angle of the recoil
electron) variables were obtained for the following conditions: all
particles are unpolarized, the photon beam is linearly polarized and
circularly polarized photon beam interacts with a polarized electron
target. In Section 4 the similar calculations were done for the
$\gamma -e$ diagrams. Section 5 contains the results of calculation
of the double (over the $q^2$ and $Q^2$ variables) and single (over
the $q^2$ or $Q^2$ variable) distributions for the same polarization
situations. The results and discussion are presented in Section 6.
Finally, in Section 7 we give some conclusions.

\section{Matrix element and cross section}
\hspace{0.7cm}

The reaction of the triplet production by a photon beam on an
electron target
\begin{equation}\label{1}
\gamma (k)+ e^-(p) \to e^-(p_1) + e^+(p_3) + e^-(p_2),
\end{equation}
where 4-momenta of the particles are shown in the parentheses, is
described, in general, by eight Feynman diagrams. In the
calculations we take into account four Feynman diagrams: the
so-called Borsellino diagrams (see Fig. 1a) and the $\gamma -e$ ones
(see Fig. 1b), i.e., we neglect the effects of the identity of the
final electrons.

\begin{figure}[t]
 % \centering
\includegraphics[width=0.47\textwidth]{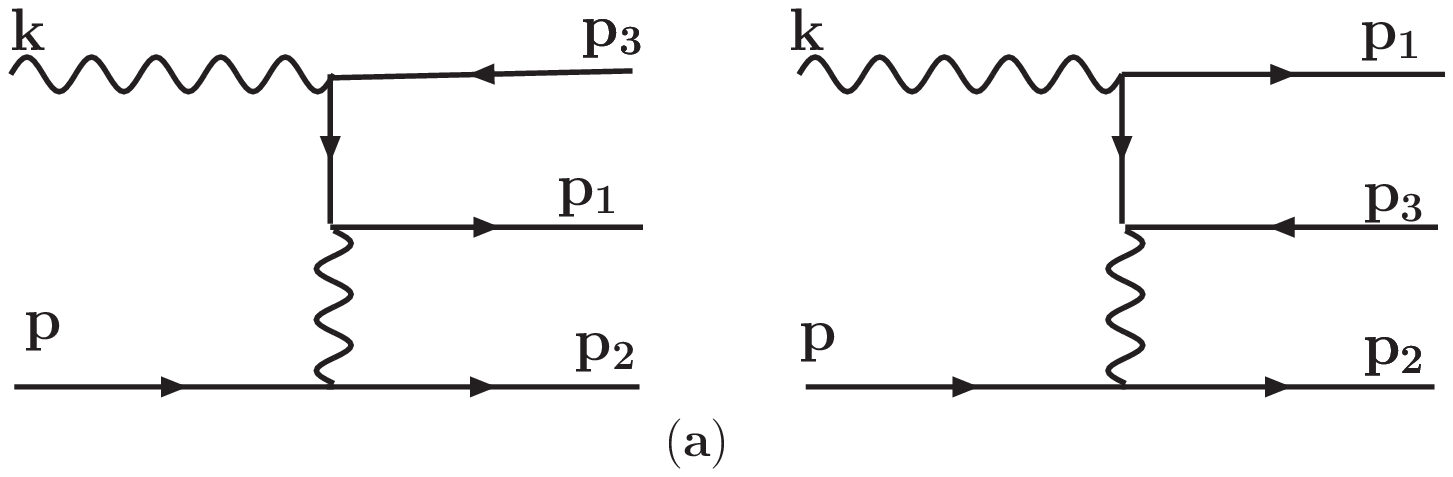}
\hfill
\includegraphics[width=0.47\textwidth]{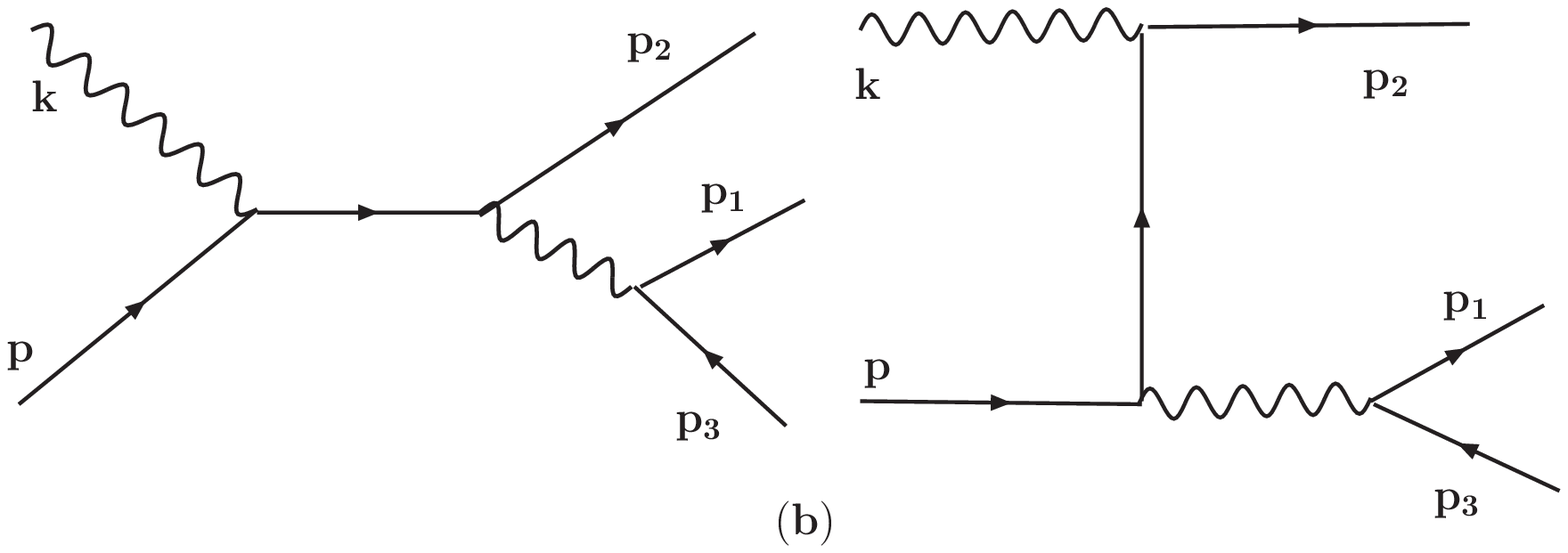}
 \parbox[t]{1\textwidth}{\caption{Feynman diagrams describing the process of the triplet
photoproduction on free electrons (the exchange diagrams are
omitted). The diagrams (a) are the Borsellino's ones, and the
diagrams (b) are the $\gamma -e$ ones.}\label{fig.1}}
\end{figure}

Therefore, in this approximation, the matrix element of the reaction
(1) can be written as follows
\begin{equation}\label{2}
M=M(B)+M(C),
\end{equation}
where $M(B)\,(M(C))$ is the contribution of two Borsellino (($\gamma
-e$) or Compton-like) diagrams. These contributions have the following form
\begin{equation}\label{3}
M(B)=(4\pi\alpha)^{\frac{3}{2}}q^{-2}A_{\mu}j_{\mu}(B), \
M(C)=(4\pi\alpha)^{\frac{3}{2}}Q^{-2}A_{\mu}j_{\mu}(C),
\end{equation}
where $q=p-p_2, \ Q=p_1+p_3, \ \alpha =e^2/4\pi =1/137,$ $A_{\mu}$ is
the polarization 4-vector of the initial photon.

The currents corresponding to the Borsellino and the $\gamma -e$
diagrams can be written as
\begin{equation}\label{4}
j_{\mu}(B)=\bar u(p_2)\gamma_{\lambda}u(p)\bar u(p_1)\hat
Q_{\mu\lambda}u(-p_3), \ \ \hat
Q_{\mu\lambda}=\frac{1}{2d_1}\gamma_{\mu}\hat
k\gamma_{\lambda}-\frac{1}{2d_3}\gamma_{\lambda}\hat
k\gamma_{\mu}+e_{\mu}^{(31)}\gamma_{\lambda},
\end{equation}
$$j_{\mu}(C)=\bar u(p_1)\gamma_{\lambda}u(-p_3)\bar u(p_2)\hat
K_{\mu\lambda}u(p), \ \ \hat
K_{\mu\lambda}=\frac{1}{2d_2}\gamma_{\mu}\hat
k\gamma_{\lambda}+\frac{1}{2d}\gamma_{\lambda}\hat
k\gamma_{\mu}-e_{\mu}^{(20)}\gamma_{\lambda}, $$ where we introduce
the following notation
$$e_{\mu}^{(20)}
=\frac{p_{2\mu}}{d_2}-\frac{p_{\mu}}{d}, \ \ e_{\mu}^{(31)}
=\frac{p_{3\mu}}{d_3}-\frac{p_{1\mu}}{d_1}, $$ $d=(k p), \
d_i=(k p_i)$ (i=1,2,3).

In the case when the polarization state of the photon beam is
described by the spin-density matrix, the square of the matrix
element can be written as follows
\begin{equation}\label{5}
|M|^2=(4\pi\alpha)^3\rho_{\mu\nu}^{\gamma}[q^{-4}T_{\mu\nu}(B)+Q^{-4}T_{\mu\nu}(C)+
q^{-2}Q^{-2}T_{\mu\nu}(BC)],
\end{equation}
where $\rho_{\mu\nu}^{\gamma}$ is the photon-beam spin-density
matrix and we use the following covariant expression for it
\begin{equation}\label{6}
\rho_{\mu\nu}^{\gamma}=\frac{1}{2}\big(
[e_{1\mu}e_{1\nu}+e_{2\mu}e_{2\nu}]
+\xi_3[e_{1\mu}e_{1\nu}-e_{2\mu}e_{2\nu}]+
\xi_1[e_{1\mu}e_{2\nu}+e_{2\mu}e_{1\nu}]-
i\xi_2[e_{1\mu}e_{2\nu}-e_{2\mu}e_{1\nu}]\big),
\end{equation}
where $\xi_i$ are the Stokes parameters and the mutually orthogonal
space-like 4-vectors $e_1$ and $e_2,$ relative to which the photon
polarization properties are defined, have to satisfy the following
relations
\begin{equation}\label{7}
e_1^2=e_2^2=-1\,, \ (e_1k)=(e_2k)=(e_1e_2)=0\,.
\end{equation}

The first term inside the parentheses in r.h.s.~of Eq.$\,$(6)
corresponds to the events with unpolarized photon, the second and
third ones are responsible for the events with linear photon
polarization and the last one -- for the events with the circular
polarization. The Stokes parameters $\xi_1$ and $\xi_3,$ which
define the linear polarization degree of the photon, depend on the
choice of the 4-vectors $e_1$ and $e_2,$ whereas the parameter
$\xi_2$ does not depend.
%Because we want to investigate the events
%with circular photon polarization, we can choose these 4-vectors by
%the most convenient way, namely
%\begin{equation}\label{7}
%e_{1\mu}=\frac{dp_{2\mu}-d_2p_{\mu}}{N}, \ \ e_{2\mu}=\frac{<\mu
%kpp_2>}{N},
%\end{equation}
%where $N^2=2dd_2p\cdot p_2-m^2(d^2+d_2^2), <\mu
%abc>=\epsilon_{\mu\alpha\beta\gamma}a_{\alpha}b_{\beta}c_{\gamma}$.
As the 4-vectors $e_i$ we choose the following:
\begin{equation}\label{8}
e_1=(0, \vec{e}_1), \ \ e_2=(0, \vec{e}_2), \ \
\vec{e}_1\,^2=\vec{e}_2\,^2=1, \ \  (\vec{e}_1 \vec{e}_2)=0.
\end{equation}

In the laboratory system, we define a coordinate system with the $z$
axis directed along the photon-beam momentum $\vec{k}$, and the
$x(y)$ axis directed along the vector $\vec{e}_1(\vec{e}_2)$. In
this case, the recoil-electron momentum $\vec{p}_2$ is determined by
the polar and azimuthal angles $\theta $ and $\varphi $,
respectively (see Fig. 2).

\begin{figure}[t]
\centering
\includegraphics[width=0.47\textwidth]{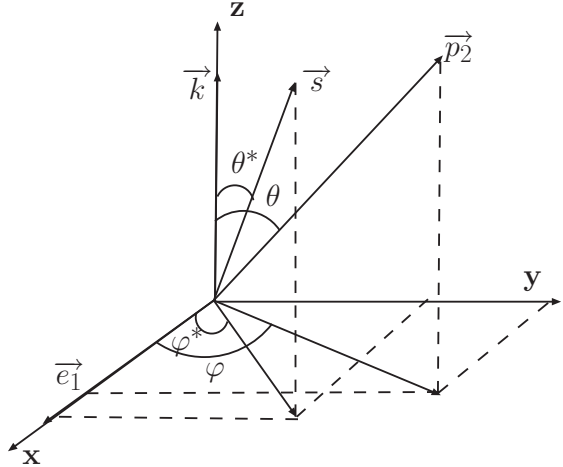}
\hfill
\parbox[t]{1\textwidth}{\caption{The angles defining the kinematics of the triplet
photoproduction process in the laboratory system. $\vec{e}_1$ is the
photon-beam polarization vector, $\vec{s}$ is the electron
polarization vector, $\vec{k}(\vec{p}_2)$ is the photon-beam
(recoil-electron) momentum.}\label{fig.2}}
\end{figure}

If we define two orthogonal space-like unit four-vectors
$$ A_{1\mu}=\frac{dp_{2\mu}-d_2p_{\mu}}{N}, \ \ A_{2\mu}=\frac{<\mu
kpp_2>}{N}, \ \ N^2=2d\,d_2(p p_2)-m^2(d^2+d_2^2), $$ for the
description of the photon polarization states, then up to the gauge
transformation we can write
$$e_{1\mu}=A_{1\mu}cos\varphi -A_{2\mu}sin\varphi , \ \
e_{2\mu}=A_{1\mu}sin\varphi +A_{2\mu}cos\varphi , $$ where $<\mu
kpp_2>=\varepsilon_{\mu\alpha\beta\gamma}k_{\alpha}p_{\beta}p_{2\gamma}, \
\varepsilon_{1234}=1$.

The tensors $T_{\mu\nu}(B) (T_{\mu\nu}(C))$ and $T_{\mu\nu}(BC)$
correspond to the contribution of the Borsellino ($\gamma -e$)
diagrams and to the interference between the Borsellino and $\gamma
-e$ diagrams. They are defined as follows
\begin{equation}\label{9}
T_{\mu\nu}(B)=j_{\mu}(B)j_{\nu}^*(B), \ \
T_{\mu\nu}(C)=j_{\mu}(C)j_{\nu}^*(C),
\end{equation}
$$T_{\mu\nu}(BC)=j_{\mu}(B)j_{\nu}^*(C)+j_{\mu}(C)j_{\nu}^*(B). $$

The differential cross section can be written in the form
\begin{equation}\label{10}
d\sigma
=\frac{(2\pi)^{-5}}{32I}|M|^2\frac{d^3p_1}{E_1}\frac{d^3p_2}{E_2}
\frac{d^3p_3}{E_3}\delta^{(4)}(k+p-p_1-p_2-p_3),
\end{equation}
where $I=(k p) $. The factors, which correspond to the averaging
over the spins of the photon beam and initial electron, are included
in $|M|^2$.

Consider the experimental setup when the produced electron-positron
pair is not detected. In this case, it is necessary to integrate over
the variables of this pair. The most convenient way to do this is to
use the method of the invariant integration. In this experimental
conditions, the contribution corresponding to the C- odd
interference of the Borsellino and the $\gamma -e$ diagrams becomes
zero. It is convenient to express the phase space of the recoil
electron in terms of the invariant variables
$$
\frac{d^3p_2}{E_2}=\frac{d(-q^2)dQ^2d\varphi}{4(k p)}=\frac{d\,\Gamma}{4(kp)}\,, \ \ d\,\Gamma=d\,\Phi\,d\varphi\,, $$ where
$q^2=(p-p_2)^2, \ Q^2=(p_1+p_3)^2=(k+q)^2$ is the square of the
invariant mass of the produced electron-positron pair, $\varphi $ is
the azimuthal angle of the recoil electron (i.e., the angle between
the photon polarization vector $\vec{e}_1$ and the plane containing
the initial-photon and recoil-electron momenta, see Fig. 2).

The limits of the integration over the $Q^2$ and $q^2$ variables are
defined by the following relations
\begin{equation}\label{11}
4m^2\leq Q^2\leq (\sqrt{W^2}-m)^2, \ \ x_-\leq -q^2\leq x_+, \ \
W^2=(k+p)^2=m^2+2m\omega ,
\end{equation}
$$x_{\pm}=\frac{d}{W^2}(2d+Q^2)-Q^2\pm \frac{d}{W^2}
\sqrt{(2d-Q^2)^2-4m^2Q^2}\,; $$
$$
4m^2\leq Q^2\leq -\frac{q^2}{m}(\omega
\sqrt{1-4\frac{m^2}{q^2}}-m-\omega )\,, \ \
 x_-(Q^2=4m^2)\leq -q^2\leq x_+(Q^2=4m^2)\,, $$
where $\omega $ is the photon-beam
energy in the laboratory system. The accessible region of the $Q^2$
and $q^2$ variables is shown in Fig. 3.

\begin{figure}[t]
\centering
\includegraphics[width=0.47\textwidth]{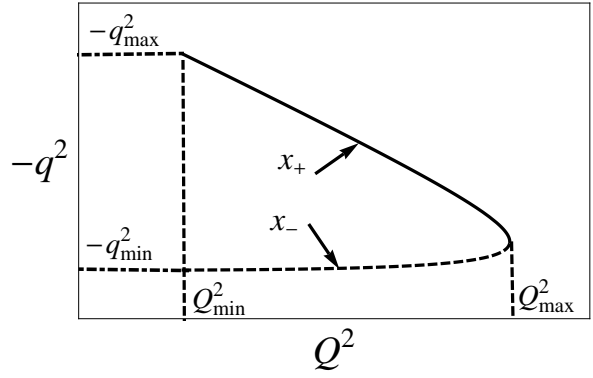}
\hfill
\parbox[t]{1\textwidth}{\caption{The allowed domain of the variables
$q^2$ and $Q^2$.}\label{fig.3}}
\end{figure}

Therefore, the distribution over the recoil-electron variables can
be written as a sum of two contributions
\begin{equation}\label{12}
\frac{d\sigma}{d\Gamma}=\frac{d\sigma (B)}{d\Gamma}+ \frac{d\sigma
(C)}{d\Gamma},
\end{equation}
where the first (second) term is
the contribution of the Borsellino ($\gamma -e$) diagrams.

\section{The Borsellino contribution}
\hspace{0.7cm}

The tensor $T_{\mu\nu}(B)$, which describes the contribution of the
Borsellino diagrams, can be represented as a product of two tensors
of the second and fourth rank
\begin{equation}\label{13}
T_{\mu\nu}(B)=t_{\lambda\rho}(B)t_{\mu\nu\lambda\rho}(B),
\end{equation}
where the current tensor $t_{\lambda\rho}(B)$ describes the
transition $e^-(p)\to\gamma^*(q)+e^-(p_2)$ and the tensor
$t_{\mu\nu\lambda\rho}(B)$ describes the process $\gamma(k)
+\gamma^*(q)\to e^-(p_1)+e^+(p_3)$.

The part of the distribution, caused by the Borsellino diagrams, can
be written as
\begin{equation}\label{14}
\frac{d\sigma (B)}{d\Gamma}=\frac{\alpha^3}{16\pi^2
d^2}\frac{1}{q^4}\Sigma (B),
\end{equation}
where we introduce
$$\Sigma
(B)=\rho_{\mu\nu}^{\gamma}Q_{\mu\nu\lambda\rho}t_{\lambda\rho}(B),
$$
\begin{equation}\label{15}
Q_{\mu\nu\lambda\rho}=\int
t_{\mu\nu\lambda\rho}(B)\frac{d^3p_1d^3p_3}{2E_12E_3}\delta^{(4)}(k+q-p_1-p_3).
\end{equation}

The Borsellino tensor $t_{\mu\nu\lambda\rho}(B)$ in the case of
unpolarized electron-positron pair was calculated in Ref.
\cite{GKLM13}. Note that the expression for this tensor in this
paper contains some misprints. Correct expression has the following
form
\begin{equation}\label{16}
t_{\mu\nu\lambda\rho}(B,0)=t_{(\mu\nu)(\lambda\rho)}(B) +
t_{[\mu\nu][\lambda\rho]}(B),
\end{equation}
where we use the index notation $(\alpha\beta)$ ($[\alpha\beta]$) to
emphasize the symmetry (antisymmetry) under permutation of the
indices $\alpha$ and $\beta.$ Note that the first tensor, which is
symmetric, contributes in the case of unpolarized or linearly
polarized photon beam. The second tensor, which is antisymmetric,
contributes in the case of a circularly polarized photon beam.
\begin{equation}\label{17}
t_{(\mu\nu)(\lambda\rho)}(B)=\frac{2}{d_1d_3}\Bigl\{g_{\lambda\rho}
[(d_1+d_3)^2g_{\mu\nu}-d_1d_3q^2e_{\mu}^{(31)}e_{\nu}^{(31)}]-2d_1d_3(1+\hat
P_{\mu\nu})g_{\lambda\nu}g_{\mu\rho}-
\end{equation}
$$-2(p_1p_3)_{\mu\nu}k_{\lambda}k_{\rho}+(1+\hat
P_{\mu\nu}+\hat P_{\lambda\rho}+\hat P_{\mu\nu}\hat
P_{\lambda\rho})g_{\rho\nu}[k_{\lambda}(d_3p_{1\mu}+d_1p_{3\mu})+
d_1d_3(p_{1\lambda}-p_{3\lambda})e_{\mu}^{(31)}]+ $$
$$+[d_3(p_1e^{(31)})_{\mu\nu}(kp_3)_{\lambda\rho}-
d_1(p_3e^{(31)})_{\mu\nu}(kp_1)_{\lambda\rho}]-
g_{\mu\nu}[(d_1+d_3)(kp_1)_{\lambda\rho}+(d_1+d_3)(kp_3)_{\lambda\rho}-
$$
$$-2(m^2+(p_1 p_3))k_{\lambda}k_{\rho}]+2d_1d_3e_{\mu}^{(31)}e_{\nu}^{(31)}
(p_1p_3)_{\lambda\rho}\Bigr\}, $$
\begin{equation}\label{18}
t_{[\mu\nu][\lambda\rho]}(B)=\frac{2}{d_1d_3}\Big\{(1-\hat{P_{\lambda\rho}})
\Big[(d_1^2+d_3^2)g_{\mu\lambda}g_{\nu\rho}+
\frac{(d_1^2p_{3\mu}-d_3^2p_{1\mu})k_{\rho}[p_1p_3]_{\mu\nu}}{d_1d_3}\Big]+
\end{equation}
$$+(1-\hat{P_{\mu\nu}}-\hat{P_{\lambda\rho}}+\hat{P_{\mu\nu}}\hat{P_{\lambda\rho}})\Big[
g_{\mu\rho}e_{\nu}^{(31)}(d_3^2p_{1\mu}-d_1^2p_{3\mu})+ $$
$$+\frac{d_1^2-(d_1-d_3)(m^2+(p_1 p_3))}{d_1}g_{\nu\alpha}k_{\rho}p_{1\mu}+
\frac{d_3^2-(d_3-d_1)(m^2+(p_1p_3))}{d_3}g_{\nu\alpha}k_{\rho}p_{3\mu}\Big]\Big\}, $$
where we use notation
$(ab)_{\lambda\rho}=a_{\lambda}b_{\rho}+b_{\lambda}a_{\rho}$ and
$[ab]_{\alpha\beta}=a_{\alpha}b_{\beta}-a_{\beta}b_{\alpha}$.

The tensor $Q_{\mu\nu\lambda\rho}$ was calculated in Ref.
\cite{BFK66} for the case of $k^2\ne 0$ using the method of the
invariant integration. The expression for this tensor in the case of
$k^2=0$ is given in Ref. \cite{BP71} and it can be written as
\begin{equation}\label{19}
Q_{\mu\nu\lambda\rho}=2\pi\beta
[D_1g_{\mu\lambda}g_{\nu\rho}+D_2g_{\lambda\rho}g_{\mu\nu}+D_3g_{\nu\lambda}g_{\mu\rho}
+\frac{D_4}{a^2}g_{\mu\nu}k_{\lambda}k_{\rho}+
\frac{D_5}{a^2}g_{\lambda\rho}q_{\mu}q_{\nu}+
\end{equation}
$$+\frac{D_6}{a^3}(g_{\lambda\mu}q_{\nu}k_{\rho}+g_{\nu\rho}q_{\mu}k_{\lambda})+
\frac{D_7}{a^3}(g_{\lambda\nu}q_{\mu}k_{\rho}+g_{\mu\rho}q_{\nu}k_{\lambda})+
\frac{D_8}{a^4}q_{\mu}q_{\nu}k_{\lambda}k_{\rho}], $$ where $\beta
=(1-4m^2/Q^2)^{1/2}$, $a=(k q)$ and the structure functions $D_1
- D_8$ have the following form
\begin{equation}\label{20}
D_1=-2-\frac{Q^4}{4a^2}+\frac{Q^2}{2a^2}(6a-m^2)+\frac{L}{a^2}[m^2(Q^2-2a-m^2)+
a(a-Q^2)],
\end{equation}
$$D_2=-\frac{Q^2}{4a^2}(3Q^2-4a+2m^2)+\frac{L}{a^2}[a(a-Q^2+2m^2)+
\frac{Q^4}{2}-m^4], $$
$$D_3=-\frac{Q^2}{4a^2}(Q^2+4a+2m^2)+\frac{L}{a^2}(a+m^2)(Q^2-a-m^2), $$
$$D_4=q^2(D_2-\frac{Q^2q^2}{a^2}+2\frac{m^2q^2}{a^2}L), \
D_5=0, \ D_6=-a^2D_1, $$
$$D_7=-a^2D_3, \ D_8=a^2(D_1+D_3), \ L=\frac{1}{\beta}ln\frac{1+\beta}{1-\beta}.
$$

\underline{Unpolarized case}.

In the case when all particles are unpolarized, we have
\begin{equation}\label{21}
\Sigma(B;0)=\frac{1}{2}(e_{1\mu}e_{1\nu}+e_{2\mu}e_{2\nu})
t_{\lambda\rho}(B;0)Q_{\mu\nu\lambda\rho}.
\end{equation}
The current tensor in the case of unpolarized initial electron is
given by
\begin{equation}\label{22}
t_{\lambda\rho}(B;0)=q^2g_{\lambda\rho}+2(pp_2)_{\lambda\rho}.
\end{equation}
Performing the contraction of the tensors in the expression (21),
one can obtain
\begin{equation}\label{23}
\Sigma(B;0)=-2\pi\beta\{2(q^2+2m^2)D_2+4dd_2\frac{1}{a^2}D_4+(D_1+D_3)[2m^2+
\frac{d^2+d_2^2}{a^2}q^2]\}.
\end{equation}
A similar expression was calculated in Ref. \cite{PV15} for the case
of the lepton pair photoproduction on a proton target, $\gamma
p\rightarrow l^+l^-p$, taking into account the Bethe-Heitler
mechanism. This process was proposed to test a lepton universality
by detecting the recoil proton.

Note that the expression (23) for the unpolarized cross section,
calculated by the authors in Ref. \cite{BP71}, is not correct. The
reason is that they used for the unpolarized photon spin-density
matrix the expression $g_{\mu\nu}/2$ and contracted it with the
tensor $Q_{\mu\nu\lambda\rho}$. It is not correct procedure because
the expression for the tensor $Q_{\mu\nu\lambda\rho}$ is not gauge
invariant (here the terms proportional to $k_{\mu}$ and $k_{\nu}$
were omitted).

\underline{Linearly polarized photon beam}.

In the case of the linearly polarized photon beam, we have
\begin{equation}\label{24}
\Sigma(B;\xi)=\frac{1}{2}[\xi_1(e_{1\mu}e_{2\nu}+e_{2\mu}e_{1\nu})+
\xi_3(e_{1\mu}e_{1\nu}-e_{2\mu}e_{2\nu})]
Q_{\mu\nu\lambda\rho}t_{\lambda\rho}(B;0).
\end{equation}
The contraction of these tensors in the above expression gives
\begin{equation}\label{25}
\Sigma(B;\xi)=-4\pi\beta\frac{1}{a^2}(D_1+D_3)(m^2a^2+dd_2q^2)
(\xi_1sin2\varphi +\xi_3cos2\varphi ).
\end{equation}

\underline{Circularly polarized photon beam}.

In the case of the circularly polarized photon beam, we have
\begin{equation}\label{26}
\Sigma(B;
S,\xi_2)=-\frac{i}{2}\xi_2(e_{1\mu}e_{2\nu}-e_{2\mu}e_{1\nu})
Q_{\mu\nu\lambda\rho}t_{\lambda\rho}(B;s).
\end{equation}
The current tensor in the case of polarized initial electron is
given by
\begin{equation}\label{27}
t_{\lambda\rho}(B;s)=2im<\lambda\rho qs>,
\end{equation}
where s is the polarization 4-vector of the initial electron which
satisfies the following conditions: $(p s)=0, \ s^2=-1$. In the
laboratory system this vector has the form: $s=(0, \vec{s})$, where
the unit vector $\vec{s}$ is determined by the polar and azimuthal
angles $\theta^*$ and $\varphi^*$, respectively.

The contraction of these tensors in the above expression gives
\begin{equation}\label{28}
\Sigma(B; s,\xi_2)=4\pi\beta m\xi_2 (D_1-D_3)[<e_1e_2qs>-\frac{1}{a}
((q e_1)<ke_2qs>-(q e_2)<ke_1qs>)].
\end{equation}
As follows from this expression, the electron polarization vector
normal to the plane $(\vec{k}, \vec{p}_2)$ does not contribute to
the polarization observable. Indeed, let us choose $s_{\mu}^N\sim
<\mu kpp_2>$ (in the laboratory system we have $s_{\mu}^N=(0,
\vec{s}^N)$, where $\vec{s}^N\sim (\vec{k}\times\vec{p}_2)$), then
we have $<e_1e_2qs^N>=0$ and $(q e_1)<ke_2qs^N>=(q e_2)<ke_1qs^N>$.
Thus, only components of the electron polarization
vector which belong to the plane $(\vec{k}, \vec{p}_2)$ give nonzero
contribution to the polarization observables, namely they are $s_z$
and the following combination $s_x\,cos\varphi +s_y\,sin\varphi $.

In the chosen coordinate system we can write
\begin{equation}\label{29}
\Sigma(B; s,\xi_2)=2\pi\beta\xi_2 (D_1-D_3)[2m|\vec{p}_2|sin\theta
(s_x\,cos\varphi +s_y\,sin\varphi)+
(q^2\frac{d+d_2}{a}+2m^2\frac{a}{d})s_z].
\end{equation}
Note that if the electron polarization vector is parallel or
antiparallel to the photon beam momentum, then the cross section is
not dependent on the azimuthal angle.

%%%%%%%%%%%%%%%%%%%%%%%%%%%%%%%%%%%%%%%%%%%%%%%%%%%%%%%%%%%%%%%%%%%
\section{The $\gamma -e$ contribution}
\hspace{0.7cm}

The tensor $T_{\mu\nu}(C)$ can be represented as a product of two
tensors of the second and fourth rank
\begin{equation}\label{30}
T_{\mu\nu}(C)=t_{\lambda\rho}(C)t_{\mu\nu\lambda\rho}(C),
\end{equation}
where the current tensor $t_{\lambda\rho}(C)$ describes the
transition $\gamma^*(Q)\to e^+(p_3)+e^-(p_1)$ and the tensor
$t_{\mu\nu\lambda\rho}(C)$ describes the process $\gamma(k)
+e^-(p)\to\gamma^*(Q)+e^-(p_2)$.

We consider the case when the produced electron-positron pair is
unpolarized. Then the current tensor is given by
\begin{equation}\label{31}
t_{\lambda\rho}(C)=-2Q^2g_{\lambda\rho}+4(p_1p_3)_{\lambda\rho}.
\end{equation}

The part of the distribution, caused by the $\gamma -e$ diagrams,
can be written as
\begin{equation}\label{32}
\frac{d\sigma (C)}{dq^2dQ^2d\varphi}=\frac{\alpha^3}{16\pi^2
d^2Q^4}\rho_{\mu\nu}^{\gamma}t_{\mu\lambda\nu\rho}(C)C_{\lambda\rho},
\end{equation}
where we introduce
\begin{equation}\label{33}
C_{\lambda\rho}=\int
t_{\lambda\rho}(C)\frac{d^3p_1d^3p_3}{2E_12E_3}\delta^{(4)}(Q-p_1-p_3).
\end{equation}

Taking into account that
$Q_{\lambda}C_{\lambda\rho}=Q_{\rho}C_{\lambda\rho}=0$ and that the
tensor $C_{\lambda\rho}$ depends only on one variable $Q_{\mu}$, we
can write the general form for the tensor $C_{\lambda\rho}$ as
\begin{equation}\label{34}
C_{\lambda\rho}=C(Q^2)(g_{\lambda\rho}-\frac{1}{Q^2}Q_{\lambda}Q_{\rho}).
\end{equation}
The function $C(Q^2)$ can be easily calculated in the c.m.s. of the
electron-positron pair. As a result we obtain
\begin{equation}\label{35}
C(Q^2)=-\frac{2\pi}{3}\beta (Q^2+2m^2).
\end{equation}

Write down the expression for the distribution,caused by the $\gamma
-e$ diagrams, in the form
\begin{equation}\label{36}
\frac{d\sigma (C)}{d\Phi}=\frac{\alpha^3}{16\pi^2
Q^4}\frac{C(Q^2)}{d^2}\Sigma(C),
\end{equation}
where
$$\Sigma(C)=\rho_{\mu\nu}^{\gamma}(g_{\lambda\rho}-\frac{1}{Q^2}Q_{\lambda}Q_{\rho})
t_{\mu\nu\lambda\rho}(C). $$

\underline{Unpolarized case}.

In the case when all particles are unpolarized, we have
\begin{equation}\label{37}
\Sigma(C;0)=\frac{1}{2}(e_{1\mu}e_{1\nu}+e_{2\mu}e_{2\nu})
(g_{\lambda\rho}-\frac{1}{Q^2}Q_{\lambda}Q_{\rho})t_{\mu\nu\lambda\rho}(C;0).
\end{equation}

The tensor $t_{\mu\nu\lambda\rho}(C)$, caused by the $\gamma -e$
diagrams, in the case of unpolarized particles has the following
expression (we present here only symmetrical (over the $\lambda,
\rho $ indices) part of this tensor since the antisymmetrical part
does not contribute in the case when the produced electron-positron
pair is unpolarized)
\begin{equation}\label{38}
t_{(\mu\nu)(\lambda\rho)}(C;0)=\frac{1}{dd_2}\Bigl\{g_{\lambda\rho}
[(d-d_2)^2g_{\mu\nu}+dd_2Q^2e_{\mu}^{(20)}e_{\nu}^{(20)}]+2dd_2(1+\hat
P_{\mu\nu})g_{\lambda\nu}g_{\mu\rho}+
\end{equation}
$$+2(pp_2)_{\mu\nu}k_{\lambda}k_{\rho}+(1+\hat
P_{\mu\nu}+\hat P_{\lambda\rho}+\hat P_{\mu\nu}\hat
P_{\lambda\rho})g_{\rho\nu}[-k_{\lambda}(dp_{2\mu}+d_2p_{\mu})+
N(p_{\lambda}+p_{2\lambda})e_{1\mu}]- $$
$$-N[\frac{1}{d_2}(p_2e_1)_{\mu\nu}(kp)_{\lambda\rho}+
\frac{1}{d}(pe_1)_{\mu\nu}(kp_2)_{\lambda\rho}]-
g_{\mu\nu}[(d_2-d)(kp_2)_{\lambda\rho}+(d-d_2)(kp)_{\lambda\rho}-
$$
$$-2(m^2-(p p_2))k_{\lambda}k_{\rho}]+2dd_2e_{\mu}^{(20)}e_{\nu}^{(20)}
(pp_2)_{\lambda\rho}\Bigr\}, $$ where $\hat P_{\mu\nu}$ is the
$\mu\leftrightarrow\nu$ permutation operator, the averaging over the
initial electron spin is taken into account in this tensor.

Performing the contraction of the tensors in the expression (37),
one can obtain
\begin{equation}\label{39}
\Sigma(C;0)=-\frac{1}{dd_2}\{2(d^2+d_2^2)+(Q^2+2m^2)[q^2+\frac{m^2}{dd_2}(d-d_2)^2]\}.
\end{equation}

\underline{Linearly polarized photon beam}.

In the case of the linearly polarized photon beam, we have
\begin{equation}\label{40}
\Sigma(C;\xi)=\frac{1}{2}[\xi_1(e_{1\mu}e_{2\nu}+e_{2\mu}e_{1\nu})+
\xi_3(e_{1\mu}e_{1\nu}-e_{2\mu}e_{2\nu})]
(g_{\lambda\rho}-\frac{1}{Q^2}Q_{\lambda}Q_{\rho})t_{\mu\nu\lambda\rho}(C;0).
\end{equation}
The contraction of these tensors in the above expression gives
\begin{equation}\label{41}
\Sigma(C;\xi)=-\frac{1}{dd_2}(Q^2+2m^2)[q^2+\frac{m^2}{dd_2}(d-d_2)^2]
(\xi_1sin2\varphi +\xi_3cos2\varphi ).
\end{equation}

\underline{Circularly polarized photon beam}.

The tensor $t_{\mu\nu\lambda\rho}(C;s)$, caused by the $\gamma -e$
diagrams, in the case of the polarized initial electron has the
following expression
\begin{equation}\label{42}
t_{\mu\nu\lambda\rho}(C;s)=\frac{im}{2d^2d_2^2}\{4dd_2(p_{\lambda}p_{\rho}+
p_{2\lambda}p_{2\rho})-
2(d^2+d_2^2)(pp_2)_{\lambda\rho}-[Q^2(d+d_2)^2-2d(d^2-
\end{equation}
$$-d_2^2)]g_{\lambda\rho}\}<\mu\nu ks>-\frac{im}{dd_2^2}<\mu\nu
kq>[-dd_2(se^{(20)})_{\lambda\rho}+(d+d_2)(p_2 s)
g_{\lambda\rho}]-\frac{im}{2d^2d_2^2}(1+\hat P_{\lambda\rho}) $$
$$<\mu\nu\rho k>[d(d+d_2)(Q^2+2d_2-2d)s_{\lambda}-
2(d(p_2 s)+d_2(k s))(d_2p_{2\lambda}-dp_{\lambda})]. $$

The contraction of this tensor with the tensor $C_{\lambda\rho}$
leads to the following expression (we omit here the terms
proportional to $k_{\mu}$ and $k_{\nu}$ since they do not contribute
to the observables)
\begin{equation}\label{43}
S_{\mu\nu}(C;s)=a_1<\mu\nu ks>+a_2<\mu\nu pk>,
\end{equation}
where we introduce
$$a_1=-im\frac{C(Q^2)}{d^2d_2^2}[4dd_2(d-d_2)-d_2Q^2(3d+d_2)-2m^2(d-d_2)^2],
$$
$$a_2=2im\frac{C(Q^2)}{d_2d^2}(d-d_2)(p_2 s). $$
From the formula (43) one can conclude that if the electron
polarization vector is perpendicular to the plane $(\vec{k},
\vec{p}_2)$ then it does not contribute to the polarization
observable. The coefficient $a_2=0$ since $(p_2s^N)=0$ and the
expression $<\mu\nu ks^N>\sim
k_{\mu}(dp_{2\nu}-d_2p_{\nu})+k_{\nu}(d_2p_{\mu}-dp_{2\mu})$ and it
also does not contribute to the polarization effects. Thus, only
components of the electron polarization vector which belong to the
plane $(\vec{k}, \vec{p}_2)$ give nonzero contribution to the
polarization observables, namely, they are $s_z$ and the following
combination $s_x\,cos\varphi +s_y\,sin\varphi.$

The contraction of this tensor with the photon spin-density matrix
leads to non-zero result in the case when the photon beam is
circularly polarized. The contraction is
\begin{equation}\label{44}
\Sigma
(C;s,\xi_2)=-\xi_2\frac{1}{dd_2^2}\Bigl\{m|\vec{p}_2|sin\theta
d_2(Q^2-q^2)(s_x\,cos\varphi +s_y\,sin\varphi
)+s_z[m|\vec{p}_2|cos\theta d_2(Q^2-q^2)+
\end{equation}
$$+2dd_2(Q^2+q^2)+2m^2(d-d_2)^2+d_2(d_2-d)Q^2]\Bigr\}, $$
$$m|\vec{p}_2|cos\theta =\frac{m^2}{2d}(Q^2-q^2)-\frac{q^2}{2}, \ \
\vec{p}_2~^2sin^2\theta =-\frac{1}{d^2}[m^2(d-d_2)^2+dd_2q^2], $$
where $s_i, i=x,y,z$ are the components of the electron polarization
vector in the laboratory system ($s_z=cos\theta^*,
s_x=sin\theta^*cos\varphi^*, s_y=sin\theta^*sin\varphi^*$). Note
that if the electron polarization vector is parallel or antiparallel
the photon beam momentum, then the cross section is not dependent on
the azimuthal angle.

%%%%%%%%%%%%%%%%%%%%%%%%%%%%%%%%%%%%%%%%%%%%%%%%%%%%%%%%%%%%%%%%%%%%
\section{Different distributions}
\hspace{0.7cm}

The differential cross section of the reaction (1), in the case when
the electron target and photon beam are polarized (the polarization
state of the photon beam is described by the Stokes parameters) and
the produced electron-positron pair is not detected, can be written as
\begin{equation}\label{45}
\frac{d\sigma }{d\Gamma}=\frac{d\sigma^{(U)}
}{d\Phi}+(\xi_1sin2\varphi +\xi_3\,cos2\varphi
)\frac{d\sigma^{(L)}}{d\Phi}+\xi_2(s_x\,cos\varphi +s_y\,sin\varphi
)\frac{d\sigma^{(CT)}}{d\Phi}+\xi_2
s_z\frac{d\sigma^{(CL)}}{d\Phi},
\end{equation}
where the first term describes the unpolarized differential cross
section of the reaction (1), the second term corresponds to the part
of the differential cross section caused by the linearly polarized
photon beam. The third (fourth) term corresponds to the part of the
differential cross section caused by the circularly polarized photon
beam and polarized initial electron in the case when the
polarization vector of the target is orthogonal (parallel) to the
photon momentum.

%%%%%%%%%%%%%%%%%%%%%%%%%%%%%%%%%%%%%%%%%%%%%%%%%%%%%%%%%%%%
The azimuthal integration in (45) leads to multiplication of the $d\,\sigma^{^{(U)}}$ and $d\,\sigma^{^{(CL)}}$ by factor 2$\pi,$ and to extract the part $d\,\sigma^{^{(CL)}}$ we have to take the difference of the events number with two opposite values of z-component of the target electron polarization, $s_z.$ To separate the part $d\,\sigma^{^{(CT)}}$ at fixed electron polarization, it is enough to take difference between the events number in the forward $(0<\varphi<\pi/2\,, \ 3\pi/2<\varphi<2\pi; \ \cos{\varphi}>0)$ and backward $(\pi/2<\varphi<3\pi/2; \ \cos{\varphi}<0)$ hemispheres. To probe the part $d\,\sigma^{^{(L)}},$ it is allowed to sum the events in the sectors
$(0<\varphi<\pi/4, \ 7\pi/4<\varphi<2\pi); \ (3\pi/4<\varphi<5\pi/4)$ and subtract the events number in the sectors
$(\pi/4<\varphi<3\pi/4); \ (5\pi/4<\varphi<7\pi/4).$

\subsection {Double differential distributions}

%%%%%%%%%%%%%%%%%%%%%%%%%%%%%%%%%%%%%%%%%%%%%%%%%%%%%%%%%%%%%

The unpolarized part of the differential cross section can be read
as
\begin{equation}\label{46}
\frac{d\sigma^{(U)}}{d\Phi}=\frac{d\sigma^{(U)}(B)}{d\Phi}+
\frac{d\sigma^{(U)}(C)}{d\Phi},
\end{equation}
\begin{equation}\label{47}
\frac{d\sigma^{(U)}(B)}{d\Phi}=\frac{\alpha^3}{2\pi
}\frac{\beta}{q^4}\frac{1}{m^2\omega^2}(A-\frac{B}{r_1^2}L),
\end{equation}
where the functions A and B are
\begin{equation}\label{48}
A=m^2+\frac{q^2}{2}+\frac{2}{r_1}(q^4+3m^2q^2+2m^4-m\omega
q^2)+\frac{2q^2}{r_1^2}[q^4+m^2(2m^2+3q^2+2\omega^2)-
\end{equation}
$$-4m\omega (m^2+2q^2)]+8m\omega (2m\omega -q^2)(m^2+2q^2)\frac{q^2}{r_1^3}+
16m^2\omega^2\frac{q^4}{r_1^4}(m^2+2q^2), $$
\begin{equation}\label{49}
B=8m^2\omega^2\frac{q^2}{r_1^2}(q^4+6m^2q^2-4m^4)+ 4m\omega
\frac{q^2}{r_1}[4m^3(m+\omega )+2mq^2(\omega -3m)-q^4]+
\end{equation}
$$+2r_1[4m^4+q^4+2m(2m-\omega )q^2]+\frac{r_1^2}{2}(q^2+2m^2)+4m(m-\omega
)q^4-8m^6+q^6-2m^2q^2(4m\omega -2\omega^2), $$
\begin{equation}\label{50}
\frac{d\sigma^{(U)}(C)}{d\Phi}=\frac{\alpha^3}{12\pi
}\frac{\beta}{Q^4}\frac{1}{m^3\omega^3}\frac{1}{r_2}(Q^2+2m^2)\{2m^2(2\omega^2-m^2)-
\frac{m}{\omega}(2\omega^2+m^2+m\omega )Q^2+
\end{equation}
$$+\frac{m}{\omega}[m^2+2\omega (m+\omega
)]q^2+\frac{m}{2\omega}Q^2q^2+\frac{1}{2}(1-\frac{m}{\omega})Q^4+
\frac{1}{2}q^4+\frac{2}{r_2}m^3\omega (Q^2+2m^2)\}, $$ where
$r_1=Q^2-q^2$ and $r_2=2m\omega +q^2-Q^2$.

The part of the differential cross section caused by the linearly
polarized photon beam can be read as
\begin{equation}\label{51}
\frac{d\sigma^{(L)}}{d\Phi}=\frac{d\sigma^{(L)}(B)}{d\Phi}+
\frac{d\sigma^{(L)}(C)}{d\Phi},
\end{equation}
\begin{equation}\label{52}
\frac{d\sigma^{(L)}(B)}{d\Phi}=\frac{\alpha^3}{\pi
}\frac{\beta}{q^4}\frac{1}{m\omega^2}\frac{1}{r_1^4}[\frac{1}{2}q^4+m^2Q^2+
2m^2(m^2-q^2)L]
\end{equation}
$$\{(m+2\omega )q^4+mQ^4+2q^2[2m\omega^2-(m+\omega )Q^2]\}, $$
\begin{equation}\label{53}
\frac{d\sigma^{(L)}(C)}{d\Phi}=\frac{\alpha^3}{12\pi
}\frac{\beta}{Q^4}\frac{1}{m^3\omega^3}\frac{1}{r_2}(Q^2+2m^2)^2[-m^2+
\frac{2\omega+m}{2\omega}q^2-\frac{m}{2\omega}Q^2+2m^3\frac{\omega}{r_2}].
\end{equation}

The part of the differential cross section, caused by the circularly
polarized photon beam and polarized initial electron in the case
when its polarization vector is orthogonal to the photon momentum,
can be written as
\begin{equation}\label{54}
\frac{d\sigma^{(CT)}}{d\Phi}=\frac{d\sigma^{(CT)}(B)}{d\Phi}+
\frac{d\sigma^{(CT)}(C)}{d\Phi},
\end{equation}
\begin{equation}\label{55}
\frac{d\sigma^{(CT)}(B)}{d\Phi}=\frac{\alpha^3}{2\pi
}\frac{\beta}{q^4}\frac{|\vec{p}_2|}{m\omega^2}\frac{1}{r_1}sin\theta
[3Q^2+q^2-(Q^2+q^2)L],
\end{equation}
\begin{equation}\label{56}
\frac{d\sigma^{(CT)}(C)}{d\Phi}=\frac{\alpha^3}{12\pi
}\frac{\beta}{Q^4}\frac{|\vec{p}_2|}{m^2\omega^3}\frac{1}{r_2}sin\theta
(Q^2-q^2)(Q^2+2m^2).
\end{equation}

The part of the differential cross section, caused by the circularly
polarized photon beam and polarized initial electron in the case
when its polarization vector is parallel to the photon momentum, can
be written as
\begin{equation}\label{57}
\frac{d\sigma^{(CL)}}{d\Phi}=\frac{d\sigma^{(CL)}(B)}{d\Phi}+
\frac{d\sigma^{(CL)}(C)}{d\Phi},
\end{equation}
\begin{equation}\label{58}
\frac{d\sigma^{(CL)}(B)}{d\Phi}=\frac{\alpha^3}{4\pi
}\frac{\beta}{r_1q^4}\frac{1}{m^2\omega^2}[3Q^2+q^2-(Q^2+q^2)L][\frac{m}{\omega}Q^2+
-q^2(1+\frac{m}{\omega}-4\frac{m\omega}{r_1})],
\end{equation}
\begin{equation}\label{59}
\frac{d\sigma^{(CL)}(C)}{d\Phi}=\frac{\alpha^3}{24\pi
}\frac{\beta}{r_2Q^4}\frac{1}{m^3\omega^3}(Q^2+2m^2)
[\frac{m}{\omega}(1+2\frac{m\omega}{r_2})(Q^2-q^2)^2+q^4-Q^4+4m\omega
(Q^2+q^2)].
\end{equation}

%%%%%%%%%%%%%%%%%%%%%%%%%%%%%%%%%%%%%%%%%%%%%%%%%%%%%%%%%%%%%%%%%%%%
\subsection{Distributions over $Q^2$ variable} \hspace{0.7cm}

Let us obtain the distributions over the square of the invariant
mass of the produced electron-positron pair, i.e., over the $Q^2$
variable. To do this, it is necessary to integrate the distributions
(46) - (59) over the $q^2$ variable. As a result, we have for the
unpolarized part of the differential cross section
\begin{equation}\label{60}
\frac{d\sigma^{(U)}}{dQ^2}=\frac{d\sigma^{(U)}(B)}{dQ^2}+
\frac{d\sigma^{(U)}(C)}{dQ^2},
\end{equation}
\begin{equation}\label{61}
\frac{d\sigma^{(U)}(B)}{dQ^2}=\frac{\alpha^3}{2\pi
}\frac{\beta}{Q^6}\frac{1}{m^2\omega^2}\Bigl
[D_1+D_2Ln(\frac{V-\Delta}{V+\Delta})+D_3Ln\Bigl
[\frac{Q^2(m+2\omega )+\omega (\Delta -V)}{Q^2(m+2\omega )-\omega
(\Delta +V)}\Bigr ]\Bigr ],
\end{equation}
$$D_1=\frac{\Delta}{3m\omega}\Bigl\{Q^6-4m\omega Q^4-8m^3\omega
Q^2+4m^2Q^2(Q^2+7\omega^2)+4m^4(Q^2+17\omega^2)+ $$
$$+\frac{L}{Q^2}\Bigl [-Q^8+m\omega Q^6+12m^3\omega
Q^4-2m^2Q^4(Q^2+8\omega^2)-16\omega m^5Q^2+ $$
$$+4m^4Q^2(Q^2-21\omega^2)+8m^6(Q^2+17\omega^2)\Bigr ]\Bigr\}, $$
$$D_2=2m\Bigl\{RQ^2(\omega -3m)-2m^3(5Q^2+4\omega^2)+\frac{L}{Q^2}\Bigl [
RQ^4(2m-\omega )+ $$
$$+2m^2RQ^2(3m-\omega )+4m^3Q^2(m^2+4m\omega
+\omega^2)-16m^5\omega^2\Bigr ]\Bigr\}, $$
$$D_3=\frac{1}{2}\Bigl [Q^4(Q^2-4m\omega
+12m^2)+8m^2Q^2(2m^2-2m\omega +\omega^2)+32m^4\omega^2- $$
$$-\frac{L}{Q^2}(Q^4+4m^2Q^2-8m^4)(Q^4-4m\omega
Q^2+4m^2Q^2+8m^2\omega^2)\Bigr ], $$
\begin{equation}\label{62}
\frac{d\sigma^{(U)}(C)}{dQ^2}=\frac{\alpha^3}{12\pi
}\frac{\beta}{Q^4}\frac{Q^2+2m^2}{m^3\omega^3}\Bigl\{
\frac{\Delta}{(m+2\omega )^2}[2m\omega
(4Q^2+\omega^2)+2m^2(Q^2+9\omega^2)+
\end{equation}
$$+16m^3\omega +4m^4+7Q^2\omega^2]-[R(Q^2+2m^2)+2m^2\omega^2]
Ln(\frac{R+\Delta}{R-\Delta})\Bigr\}, $$ where we introduce the
following notation
$$\Delta =\sqrt{(Q^2-2m\omega )^2-4m^2Q^2}, \ \ R=Q^2-2m\omega
-2m^2, \ \ V=Q^2+2m\omega\,. $$

The $Q^2$ distribution of the differential cross section caused by
the linearly polarized photon beam can be read as
\begin{equation}\label{63}
\frac{d\sigma^{(L)}}{dQ^2}=\frac{d\sigma^{(L)}(B)}{dQ^2}+
\frac{d\sigma^{(L)}(C)}{dQ^2},
\end{equation}
\begin{equation}\label{64}
\frac{d\sigma^{(L)}(B)}{dQ^2}=\frac{\alpha^3}{\pi
}\frac{\beta}{Q^6}\frac{1}{m\omega^2}\Bigl\{\frac{\Delta}{6\omega}\Bigl
[Q^4(L-\frac{Q^2}{4m^2})+2m(\frac{Q^2}{2m^2}+L)(\omega
Q^2+2m^3(1+17\frac{\omega^2}{Q^2})-4\omega m^2)-
\end{equation}
$$-m^2L(5Q^2+8\omega^2)+\frac{Q^2}{2}(Q^2-2\omega^2)
\Bigr ]+2m^2\Bigl [R(\omega +m(L-1))-2m^3+ $$ $$+4m^2\omega
L(1-\frac{m\omega}{Q^2})\Bigr ] Ln\Bigl [\frac{Q^2-2m\omega
-\Delta}{Q^2-2m\omega +\Delta}\Bigr ]\Bigr\}, $$
\begin{equation}\label{65}
\frac{d\sigma^{(L)}(C)}{dQ^2}=\frac{\alpha^3}{12\pi
}\frac{\beta}{Q^4}\frac{(Q^2+2m^2)^2}{m^3\omega^3}(2\Delta
-RLn(\frac{R+\Delta}{R-\Delta}))\,.
\end{equation}

The $Q^2$ distribution of the differential cross section, caused by
the circularly polarized photon beam and polarized initial electron
in the case when its polarization vector is orthogonal to the photon
momentum, can be written as
\begin{equation}\label{66}
\frac{d\sigma^{(CT)}}{dQ^2}=\frac{d\sigma^{(CT)}(B)}{dQ^2}+
\frac{d\sigma^{(CT)}(C)}{dQ^2},
\end{equation}
\begin{equation}\label{67}
\frac{d\sigma^{(CT)}(B)}{dQ^2}=\frac{\alpha^3\beta}{4Q^2}\frac{1}{m^2\omega^3}
\Bigl\{Q^2(1-L)\sqrt{m^2+2m\omega}+Q^2[m-3\omega +(\omega -m)L]+
\end{equation}
$$+2m\omega [\omega (3-L)+2\sqrt{Q^2}(L-2)]\Bigr\}, $$
\begin{equation}\label{68}
\frac{d\sigma^{(CT)}(C)}{dQ^2}=-\frac{\alpha^3}{48
}\frac{\beta}{Q^4}\frac{(Q^2+2m^2)}{m^2\omega^2}\Bigl\{8m^2+
\frac{1}{\sqrt{m}}(m+2\omega)^{-3/2}[\Delta^2+4mR(m+2\omega)
]\Bigr\}\,.
\end{equation}

The $Q^2$ distribution of the differential cross section, caused by
the circularly polarized photon beam and polarized initial electron
in the case when its polarization vector is parallel to the photon
momentum, can be written as
\begin{equation}\label{69}
\frac{d\sigma^{(CL)}}{dQ^2}=\frac{d\sigma^{(CL)}(B)}{dQ^2}+
\frac{d\sigma^{(CL)}(C)}{dQ^2},
\end{equation}
\begin{equation}\label{70}
\frac{d\sigma^{(CL)}(B)}{dQ^2}=\frac{\alpha^3}{4\pi
}\frac{\beta}{Q^2}\frac{1}{m^2\omega^2}\Bigl\{2\Delta
(7-3L)+2[2(Q^2-3m\omega )-L(Q^2-2m\omega
)]Ln(\frac{V-\Delta}{V+\Delta})+
\end{equation}
$$+\frac{1}{\omega}[\omega (L-3)(Q^2-4m\omega )+mQ^2(1-L)]Ln\Bigl [\frac{Q^2(m+\omega
)-2m\omega^2+\omega\Delta}{Q^2(m+\omega
)-2m\omega^2-\omega\Delta}\Bigr ]\Bigr\}, $$
\begin{equation}\label{71}
\frac{d\sigma^{(CL)}(C)}{dQ^2}=\frac{\alpha^3}{12\pi
}\frac{\beta}{Q^4}\frac{Q^2+2m^2}{m^3\omega^2}
\Bigl\{\frac{\Delta}{(m+2\omega )^2}[(m+3\omega )R+4m(m+2\omega
)^2]+
\end{equation}
$$+2m(m\omega +m^2-Q^2)Ln(\frac{R+\Delta}{R-\Delta})\Bigr\}. $$

%%%%%%%%%%%%%%%%%%%%%%%%%%%%%%%%%%%%%%%%%%%%%%%%%%%%%%%%%%%%%%%%%%%%
\subsection{Distributions over $q^2$ variable} \hspace{0.7cm}

Let us obtain the distributions over the $q^2$ variable. To do this,
it is necessary to integrate the distributions (46) - (59) over the
$Q^2$ variable. As a result, we have for the unpolarized part of the
differential cross section
\begin{equation}\label{72}
\frac{d\sigma^{(U)}}{dq^2_m}=\frac{d\sigma^{(U)}(B)}{dq^2_m}+
\frac{d\sigma^{(U)}(C)}{dq^2_m}\,, \ \ q^2_m=-q^2\,,
\end{equation}
\begin{equation}\label{73}
\frac{d\sigma^{(U)}(B)}{dq^2_m}=\frac{\alpha^3}{2\pi
}\frac{1}{m^2\omega^2}\Bigl\{\frac{1}{2q^4}(q^4-2m\omega
q^2+4m^2q^2+4m^4)Ln(\frac{1+r}{1-r})Ln\Bigl
[\frac{m^3Y}{q^2\omega^2}(1-\bar\beta )^{-2}\Bigr ]+
\end{equation}
$$+\frac{1}{q^4}[2q^4-2m\omega
q^2+5m^2q^2+2m^4]LnG_1+\frac{2}{q^4\bar\beta^5}Ln\Bigl
[\frac{2m^2(Y+m)-Yq^2(1-r\bar\beta )}{2m^2(Y-m)}\Bigr ] $$
$$\Bigl [8\frac{m^7}{q^4}(\omega -4m-2\frac{m}{q^2}(4m^2+2m\omega
-\omega^2)-4m^3\frac{\omega^2}{q^4})+4\frac{m^4}{q^2}(\omega^2-8m\omega
+11m^2)+ $$
$$+q^2(q^2-m\omega -7m^2)+4m^3(3\omega +m)\Bigr
]+2\bar\beta^{-5}ArcTanh(r\bar\beta^{-1})\Bigl
[1-64\frac{\omega^2m^{10}}{q^{12}}- $$
$$-2m\frac{\omega +2m}{q^2}-\frac{8}{3}\frac{m^4}{q^6}(4\omega^2+24m\omega
-21m^2)+\frac{32}{3}\frac{m^8}{q^{10}}(7\omega^2-6m\omega -12m^2)+
$$ $$+\frac{8}{3}\frac{m^2}{q^4}(\omega^2+9m\omega
-6m^2)+\frac{16}{3}\frac{m^6}{q^8}(4\omega^2+3m\omega +12m^2)\Bigr
]-r\Bigl [2-\frac{Y}{m}+ $$ $$+2\frac{m}{q^2}(3m-3\omega
-Y)+\frac{4}{3}\frac{m^2}{q^4}(3m^2-\omega^2)+
\frac{16}{3}\frac{m^5(m^2+2q^2)}{\omega q^6}(1-\bar\beta)^{-3}- $$
$$-\frac{4}{3}\frac{m^3}{\omega
q^6}(1-\bar\beta)^{-2}(q^2(6q^2-11m\omega +3m^2)-4\omega
m^3)+\frac{2}{3}\frac{m}{\omega q^4}(1-\bar\beta)^{-1}(3q^4- $$
$$-21m\omega q^2+m^2(4\omega^2+9q^2-6m\omega +6m^2))\Bigr ]-Ln(\frac{1+r}{1-r})
\Bigl [\frac{1}{2mq^4}(q^2+2m^2)(Yq^2- $$
$$-2m^3)-\frac{8}{3}\frac{m^5}{\omega
q^8}(1-\bar\beta)^{-3}(q^4+6m^2q^2 -4m^4)+2\frac{m^3}{\omega
q^6}(1-\bar\beta)^{-2}(q^2(q^2-2m\omega +6m^2)- $$
$$-4m^3(\omega +m))+\frac{m}{\omega
q^6}(1-\bar\beta)^{-1}(-q^6+4m\omega q^4+4m^2q^2(2m\omega
-q^2-\omega^2)+8m^6)\Bigr ]- $$
$$-\frac{1}{q^4}(q^4-2m\omega q^2+4m^2q^2+4m^4)\Bigl
[Li_2(\frac{1-r}{2})-
Li_2(\frac{1-r}{1-\bar\beta})-Li_2(\frac{1-r}{1+\bar\beta})+ $$
$$+Li_2(\frac{1+r}{1+\bar\beta})+
Li_2(\frac{1+r}{1-\bar\beta})-Li_2(\frac{1+r}{2})\Bigr ]\Bigr \},
$$
\begin{equation}\label{74}
\frac{d\sigma^{(U)}(C)}{dq^2_m}=\frac{\alpha^3}{12\pi
}\frac{1}{m^3\omega^3}\Bigl\{\frac{r}{6m\omega y^3}\Bigl
[48\omega^2m^8+96\omega
m^7(\omega^2-q^2)+48m^6(3\omega^2q^2+\omega^4)+
\end{equation}
$$+4m^5(26\omega^3q^2+40\omega^5+23\omega q^4)+2m^4q^2(11q^4+40\omega^2q^2
+120\omega^4)+2\omega q^2m^3(12\omega^4+ $$
$$+70q^2\omega^2+7q^4)+q^4m^2(-3q^4+40\omega^2q^2
+36\omega^4)+m\omega q^6(5q^2+18\omega^2)+3\omega^2q^8+ $$
$$+3\bar\beta \omega q^2y^3(\omega -m)+12y\omega m^5\frac{(y+2m^2)^2}{2m^2+
(\bar\beta -1)q^2}\Bigr ]-\frac{2}{3}\frac{rm^3}{\omega
Yq^2y^2}\Bigl [\omega q^4(q^2+6m\omega )+ $$
$$+4\omega m^2q^2(q^2+4\omega^2)+2m^3(q^4+4q^2\omega^2+8\omega^4)\Bigr
]+\frac{1}{2\omega}[2m(m+\omega )^2-\omega q^2]LnG_1- $$
$$-\frac{LnG_2}{y^3\sqrt{y(y-4m^2)}}\Bigl [16\omega m^7(m-\omega
)^2-32q^2m^7(m+\omega )+4m^4q^4(2m^2+3q^2)-8m^4\omega^4(5q^2-
$$
$$-6m^2)+4m^4\omega^2q^2(12m^2+13q^2)+16m^5\omega^3(5q^2-\omega^2)+4\omega
q^4m^3(11m^2-11\omega^2+4q^2)- $$
$$-q^8(8m\omega +q^2)-2m^2q^6(13\omega^2-q^2)\Bigr ]\Bigr\}, $$
where
$$\bar\beta^2=1-\frac{4m^2}{q^2}, \ Y=m+\omega (1-\bar\beta ), \
r=\sqrt{1-\frac{4m^3}{Yq^2}}, \ y=q^2+2m\omega , \
G_1=\frac{q^2}{2m^3}(1+r)Y-1, $$
$$G_2=\frac{1}{2m^2\omega}[2m^2+q^2(\bar\beta -1)]^{-1}\{q^2Y
[r\sqrt{y(y-4m^2)}+y-2m^2]-2ym^3\}\,. $$

The $q^2$ distribution of the differential cross section caused by
the linearly polarized photon beam can be read as
\begin{equation}\label{75}
\frac{d\sigma^{(L)}}{dq^2_m}=\frac{d\sigma^{(L)}(B)}{dq^2_m}+
\frac{d\sigma^{(L)}(C)}{dq^2_m},
\end{equation}
\begin{equation}\label{76}
\frac{d\sigma^{(L)}(B)}{dq^2_m}=\frac{\alpha^3}{\pi
}\frac{1}{m\omega^2q^4}\Bigl\{-\frac{r}{6}\Bigl [3q^2(m+\omega
)+4m\omega^2-3\frac{m^2\bar\beta}{\omega}(1-\bar\beta
)^{-2}(q^2+2mY)+
\end{equation}
$$+8\frac{m^5}{q^2}(1-\bar\beta
)^{-2}+4\frac{m^4}{\omega q^2}(1-\bar\beta )^{-3}(q^2+2m^2)\Bigr
]+m^3LnG_1+ $$
$$+2\frac{m^4(m^2-q^2)}{\omega q^2}(1-\bar\beta
)^{-2}[\frac{8}{3}\frac{m^2}{q^2}(1-\bar\beta )^{-1}-1
]Ln(\frac{1+r}{1-r})+ $$
$$+\frac{4}{3}\frac{\omega^2m^3}{\bar\beta q^2}
Ln\Bigl [\frac{2m^2(Y+m)-Yq^2(1-r\bar\beta )}{2m^2\omega
}(1-\bar\beta )^{-1}\Bigr ]\Bigr \}, $$
\begin{equation}\label{77}
\frac{d\sigma^{(L)}(C)}{dq^2_m}=\frac{\alpha^3}{12\pi
}\frac{1}{m^3\omega^3}\Bigl\{-\frac{2m^2r}{6\omega y^3}\Bigl
[24\omega^2m^5+48\omega
m^4(\omega^2-q^2)+24\omega^2m^3(\omega^2+3q^2)+
\end{equation}
$$+2\omega q^2m^2(23q^2+56\omega^2)+mq^4(11q^2+94\omega^2)+22\omega
q^6-\frac{3q^2y^3}{2m^2}(m+\omega (1-\bar\beta))+ $$
$$+4ym^3\frac{mq^2+2y\omega}{m+\omega (1-\bar\beta)}-
6y\omega m^2\frac{(y+2m^2)^2}{2m^2+q^2(\bar\beta-1)}\Bigr
]+\frac{1}{\omega}(m^3+2m^2\omega -\omega q^2)LnG_1- $$
$$-\frac{LnG_2}{y^3\sqrt{y(y-4m^2)}}\Bigl [16\omega
m^9-32m^8(q^2+\omega^2)-16\omega
m^7(2q^2+\omega^2)-8m^6(q^4-4\omega^2q^2-4\omega^4)+ $$
$$+4\omega
q^2m^5(11q^2+16\omega^2)-4m^4q^2(4\omega^4-12\omega^2q^2-3q^4)-
16\omega m^3q^4(2\omega^2-q^2)- $$
$$-2m^2q^6(12\omega^2-q^2)-8m\omega q^8-q^{10}\Bigr ]\Bigr\}\,. $$

The $q^2$ distribution of the differential cross section, caused by
the circularly polarized photon beam and polarized initial electron
in the case when its polarization vector is orthogonal to the photon
momentum, can be written as
\begin{equation}\label{78}
\frac{d\sigma^{(CT)}}{dq^2_m}=\frac{d\sigma^{(CT)}(B)}{dq^2_m}+
\frac{d\sigma^{(CT)}(C)}{dq^2_m},
\end{equation}
\begin{equation}\label{79}
\frac{d\sigma^{(CT)}(C)}{dq^2_m}=\frac{\alpha^3}{12\pi
}\frac{1}{m^2\omega^3}\frac{1}{\sqrt{b(a+c)}}\Bigl\{\frac{\bar\beta
q^2}{3my}\frac{a+c}{a}(3ay+4m^2q^2)K\Bigl
(\frac{a(b-c)}{b(a+c)}\Bigr )+
\end{equation}
$$+\frac{2m^2}{3\omega y^2}\frac{a+c}{ac}E\Bigl
(\frac{a(b-c)}{b(a+c)}\Bigr )[3ab(3q^4+4m\omega (m\omega
-2m^2+2q^2))-yq^2(a(b+c)-bc)]- $$
$$-4\frac{\omega\bar\beta}{y^2}q^2(2m^2+y)(4m^2-y)\Pi \Bigl
(\frac{(a+y)(b-c)}{(a+c)(b-y)}\mid \frac{a(b-c)}{b(a+c)}\Bigr )+ $$
$$+2\frac{\bar\beta}{m^2}q^2[(m+\omega )q^2-2m^2\omega ]\Pi \Bigl
(\frac{(c-b)}{(a+c)}\mid \frac{a(b-c)}{b(a+c)}\Bigr ) \Bigr\}, $$
where K, E, and $\Pi$ are the standard elliptic functions
\cite{GR-TI} and
$$a=-\frac{q^2}{m}(m+\omega +\omega\bar\beta ), \
b=\frac{q^2}{m}(m+\omega -\omega\bar\beta ), \ c=4m^2. $$ Note that
we can not represent the analytic expression for the contribution
$d\sigma^{(CT)}(B)/dq^2_m$ in terms of the elementary or known special
functions.

The $q^2$ distribution of the differential cross section, caused by
the circularly polarized photon beam and polarized initial electron
in the case when its polarization vector is parallel to the photon
momentum, can be written as
\begin{equation}\label{80}
\frac{d\sigma^{(CL)}}{dq^2_m}=\frac{d\sigma^{(CL)}(B)}{dq^2_m}+
\frac{d\sigma^{(CL)}(C)}{dq^2_m},
\end{equation}
\begin{equation}\label{81}
\frac{d\sigma^{(CL)}(B)}{dq^2_m}=\frac{\alpha^3}{4\pi
}\frac{1}{m^2\omega^2q^4}\Bigl\{\frac{rYq^2}{4m\omega}[q^2(7Y+8m-16\omega
)-6m^3]-8q^2\bar\beta (m\omega -q^2)ArcTanh\frac{r}{(1-\bar\beta)}+
\end{equation}
$$+8q^2\frac{mY}{1-\bar\beta}(Ln(\frac{1+r}{1-r})-2r)-Ln(\frac{1+r}{1-r})\Bigl
[\frac{m^2}{\omega}(3m^3-4\omega
q^2)+\frac{Yq^4}{2m\omega}(Y+2m-2\omega )+ $$
$$+4m\omega q^2+q^2(q^2-2m\omega
)(Ln\Bigl [\frac{\omega^2q^2(1-\bar\beta)^2}{m^3Y}\Bigr ]+4)\Bigr
]+2q^2(q^2-2m\omega )\Bigl [Li_2(\frac{1+r}{1-\bar\beta})- $$
$$-Li_2(\frac{1-r}{1-\bar\beta})+Li_2(\frac{1+r}{1+\bar\beta})-
Li_2(\frac{1-r}{1+\bar\beta})+Li_2(\frac{1-r}{2})-Li_2(\frac{1+r}{2})\Bigr
]\Bigr \}, $$
\begin{equation}\label{82}
\frac{d\sigma^{(CL)}(C)}{dq^2_m}=\frac{\alpha^3}{24\pi
}\frac{1}{m^3\omega^3}\Bigl\{\frac{1}{\omega}[q^2(\omega
+m)-2m\omega^2]LnG_1+4\frac{m\omega}{\sqrt{y(y-4m^2)}}\Bigl
[q^2+m(\omega -3m)-
\end{equation}
$$-40\frac{\omega m^7}{y^3}+4\frac{m^5}{y^2}(\omega
+2m)+2\frac{m^3}{y}(2\omega -3m)\Bigr ]Ln\Bigl [\frac{\omega
G_2[2m^2+q^2(\bar\beta -1)]}{Yq^2-my} \Bigr ]- $$
$$-r\Bigl [\frac{\omega -m}{m\omega}(4m^3-Yq^2)+80\frac{\omega^2
m^6}{y^3}+\frac{16}{3}\frac{\omega m^4}{y^2}(\omega
-3m)-\frac{4}{3}\frac{\omega m^2}{y}(5\omega -7m)+ $$
$$+\frac{5}{3}\frac{1}{\omega}(y\omega +mq^2)+
8\frac{\omega^2m^5}{y^2}\frac{y+2m^2}{Yq^2-my}+\frac{4}{3}\frac{m^3}{Y}(1+
\frac{m}{\omega}+2\frac{m\omega}{y^2}(y-2m^2))\Bigr ]\Bigr \}. $$

%%%%%%%%%%%%%%%%%%%%%%%%%%

\section{Results and discussion}

There are a few reasons to investigate process of the triplet photoproduction in the
framework of a more elaborative approach.

Our analysis of the triplet production is caused mainly by the search for a physics beyond SM in the frame of the project IRIDE \cite{IRIDE}. It is assumed that there is a new light particle (U-boson \cite{FAP} which is one  of the possible dark-matter candidate) that does not interact with the matter fields of the SM but can mix with a photon.

In the last time there were a number of the
experiments on the measuring of the electron-positron invariant mass
distributions in different processes. The inclusive dielectron
spectra, measured by the HADES (Darmstadt) in the collisions of the 3.5 GeV proton with the hydrogen, niobium
and other targets, were presented in the paper \cite{HADES}. The mass
range M(U)=20 - 550 MeV has been investigated. The results of a
search for a dark photon in the reaction $e^+e^-\to \gamma U, U\to
e^+e^-, \mu^+\mu^-$ using the BABAR detector were given in
\cite{L14}. The dark photon masses in the range 0.02 - 10.2 GeV were
investigated. Strong constraints on sub-GeV dark photon from SLAC
beam-dump experiment were given in the paper \cite{B14}. Similar
experiments were performed at KEK and Orsay \cite{A12}. The physics
motivation for a search of the dark photon at JLab are presented in
\cite{E14}. This is not a complete list of the current and planned
experiments to search for a dark photon. A review of the theoretical and experimental activity related to the
search for the particles in various scenarios of the physics beyond
the SM can be found in Ref.~\cite{S15}.

We think that the measurement of the distribution over the invariant mass
of the created electron-positron pair in the process (1)
would be a good method to search for a {\bf light} dark photon.
The contribution of the Borsellino diagrams is a background in search of this effect, so, it is necessary to separate the contribution of the Compton-like diagrams (Fig.~1~(b)) when the scattered virtual photon converts into the electron-positron pair
and where
the signal from the dark photon may be measured. Because this contribution, in contrast to the Borsellino one, decreases with the growth of the photon-beam energy, it is reasonable, in such investigation, to restrict yourselves to the low and intermediate photon-beam energies. Besides, one has to have the precise knowledge of the
background due to the double-photon $e^+e^-$-system (in fact the contribution of the Borsellino diagrams) and to try to find the kinematic regions where this background is smaller or the same order as the Compton contribution.

The next reason is the investigation of the possibility of determining the circular
polarization degree of a high energy photon by measuring the
asymmetry in the triplet production by a circularly polarized photon
beam on a polarized electron target. It is necessary also to analyze the influence of the
($\gamma-e^-$) diagrams contribution on the calculated observables.

The authors of Ref.~\cite{BP71} calculated the cross section
asymmetry caused by the azimuthal angle of the recoil electrons in the reaction (1) for the case of the
linearly polarized photons. They suggested to
use this effect for the analysis of the photon beam polarization. The
authors took into account only the Borsellino diagrams, and, up to now, there are no calculations beyond this approach. We consider
the influence of the ($\gamma-e^-$) diagrams contribution on the
asymmetry as a function of the $q^2_m$ variable and calculate its dependence on the $Q^2$ variable.

%Note that we neglect by the identity of the final electrons. It means that we clearly distinguish between created and recoil electrons.
%If not, the analyse is not so simple and requires more theoretical calculations, and to catch the effects caused by the U-boson we have %(on our opinion) with the necessity to study the double-differential distribution over invariants $(p_1+p_3)^2$ and $(p_2+p_3)^2.$

In Fig.~4 we present the integrated over $Q^2$ and $q^2_m$ (in the limits (11)) parts of the cross section, given in the r.h.s.~of Eq.~(45), and corresponding polarization asymmetries.
%%%%%%%%%%%%%%%%%%Mer%%%%%%%%%%%%%%%%%%%%%
They are obtained by the numerical integration over $Q^2$ of our analytic distributions, derived in subsection 5.2.
%%%%%%%%%%%%%%%%%%%%%%%%%%%%%%%%%%%%%%%%
For unpolarized case we show also the well known asymptotic cross section (at large photon-beam energy $\omega\gg m$), caused by the Borsellino diagrams only, which in our normalization reads \cite{AB}
\begin{equation}\label{83}
\sigma=\frac{\alpha^3}{2\pi\,m^2}\Big(\frac{28}{9}\ln\frac{2\omega}{m}-\frac{218}{27}\Big)\,.
\end{equation}

The asymmetries, as functions of the photon beam energy (Fig.~4), are defined as follows
\begin{equation}\label{84}
A^{^I}=\frac{\sigma^{^I}}{\sigma^{^U}}\,, \ \ I=L\,, \ CL\,, \ CT\,,
\end{equation}
where $\sigma^{^U}$ is unpolarized cross section caused by a sum of the Borselino and ($\gamma-e^-$) contributions.

The differential asymmetries, as functions of the $q^2_m$ or $Q^2$ variables and photon-beam energy, are defined as
\begin{equation}\label{85}
A^{^I}_{q^2_m}=\frac{d\,\sigma^{^I}/d\,q^2_m}{d\,\sigma^{^U}/d\,q^2_m}\,, \ \ A^{^I}_{Q^2}=\frac{d\,\sigma^{^I}/d\,Q^2}{d\,\sigma^{^U}/d\,Q^2}\,.
\end{equation}

\begin{figure}[t]
 \centering
\includegraphics[width=0.35\textwidth]{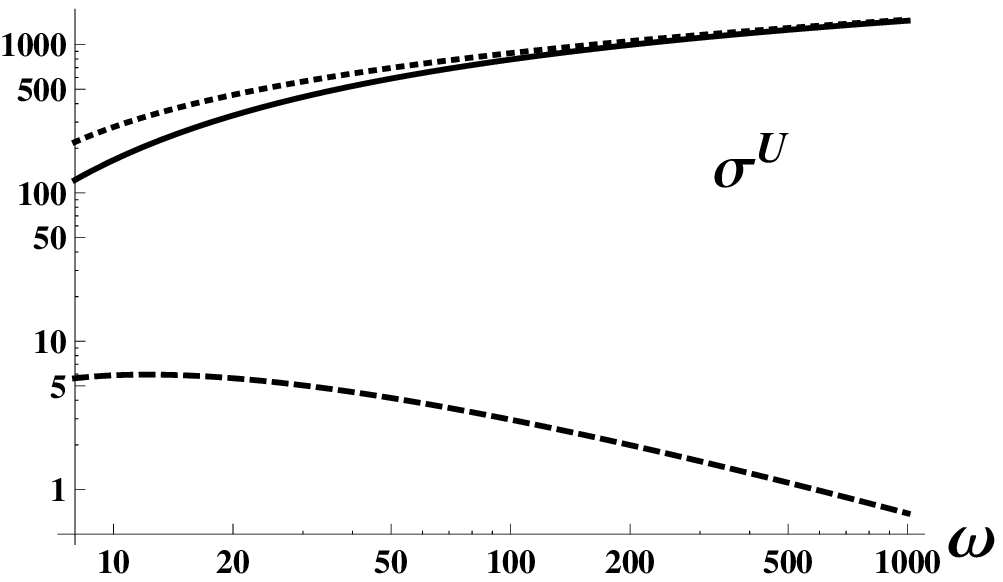}
\hspace{0.5cm}
\includegraphics[width=0.35\textwidth]{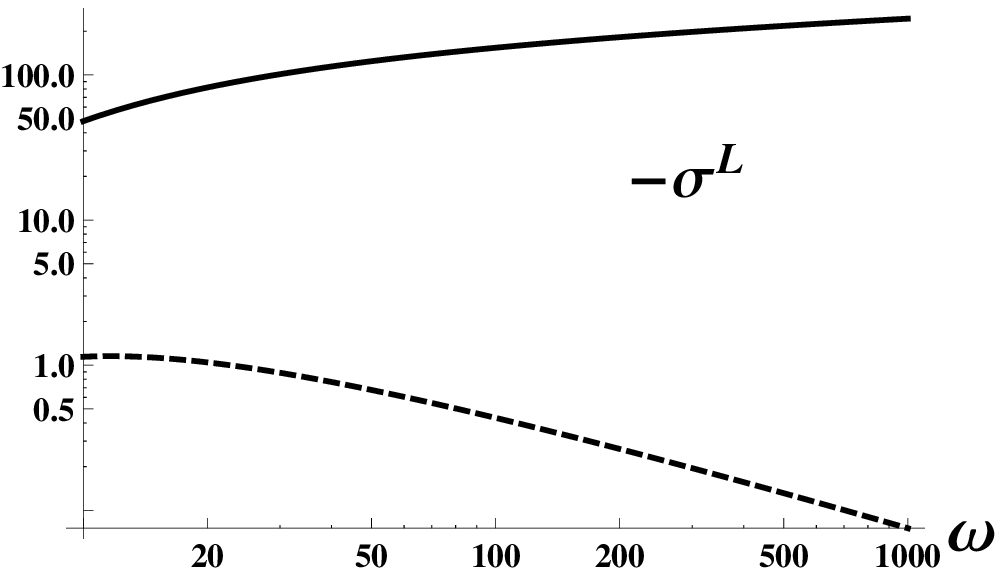}

\vspace{1cm}
\includegraphics[width=0.35\textwidth]{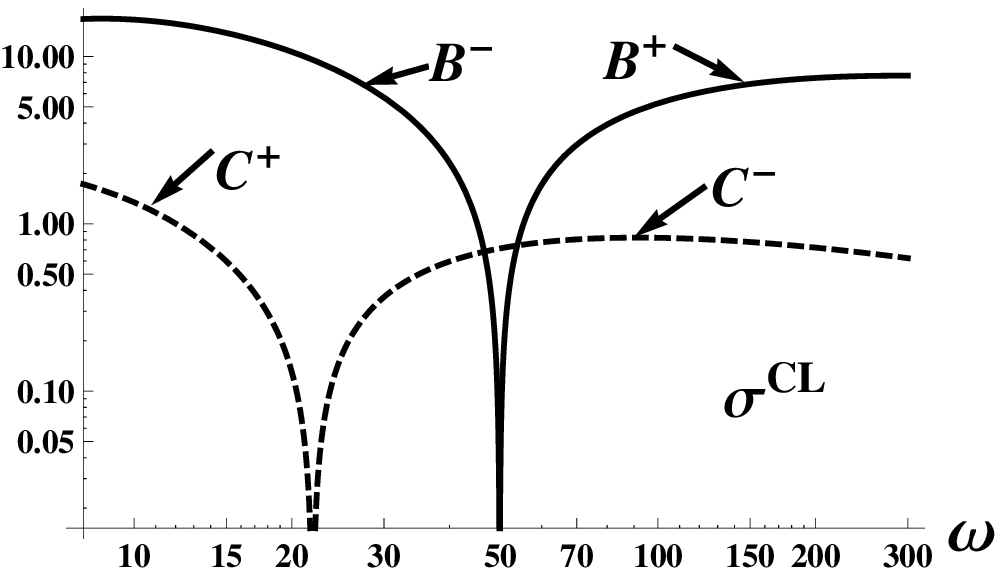}
\hspace{0.5cm}
\includegraphics[width=0.35\textwidth]{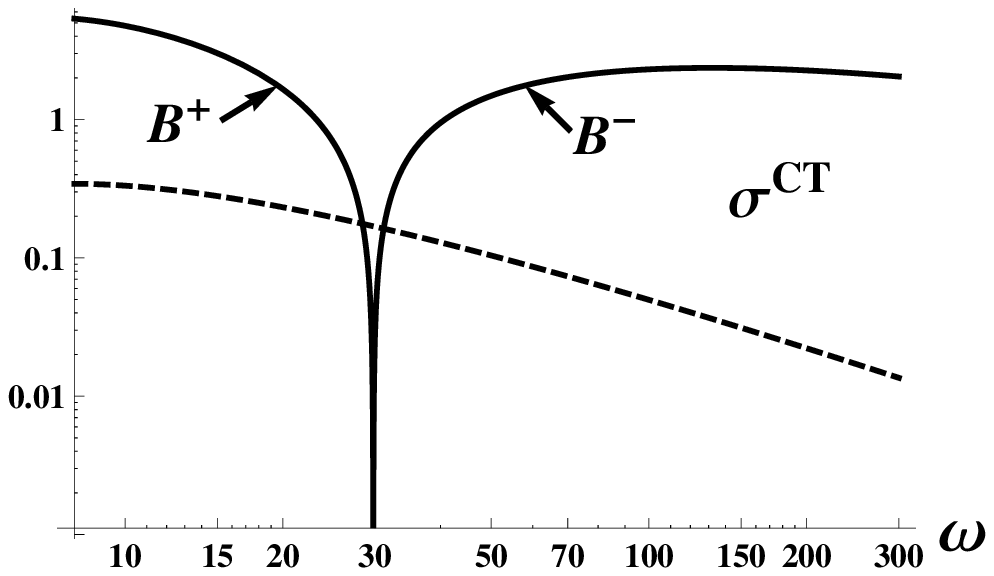}
\hspace{0.5cm}

\vspace{1cm}
\includegraphics[width=0.28\textwidth]{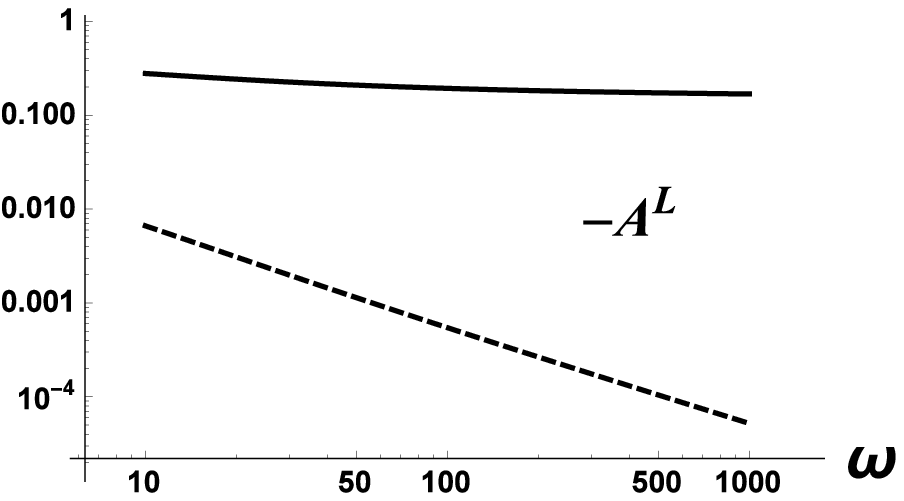}
\hspace{0.5cm}
\includegraphics[width=0.28\textwidth]{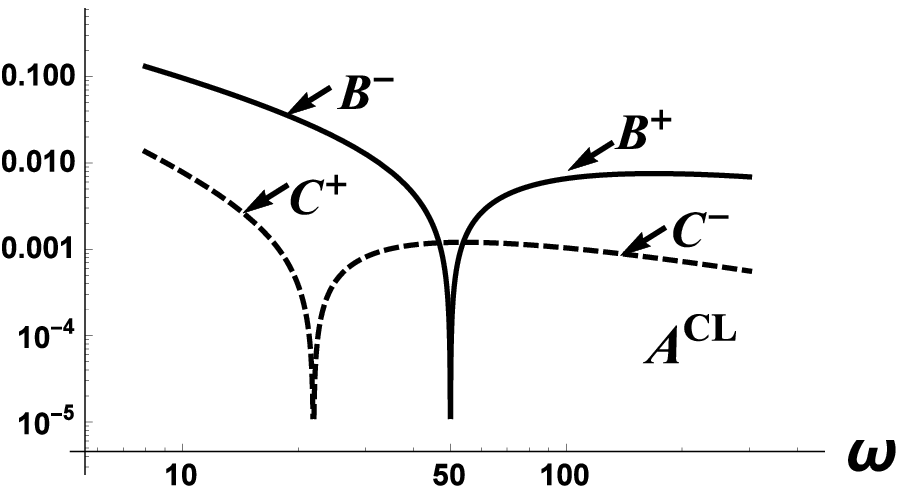}
\hspace{0.5cm}
\includegraphics[width=0.28\textwidth]{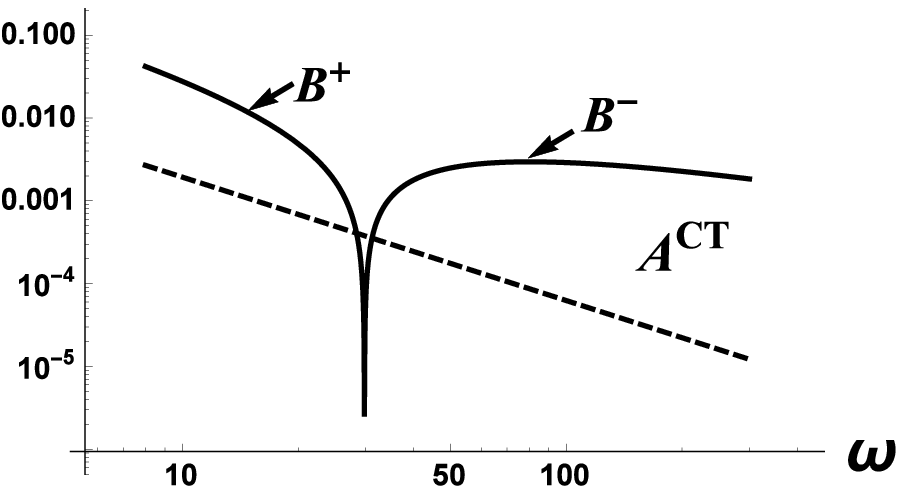}

\parbox[t]{1\textwidth}{\caption{The parts of the cross section (45) (the first and second rows) integrated numerically over all the phase space except the azimuthal angle, in
 $\mu\,b$, as a function of the photon-beam energy $\omega,$ in MeV. The solid (dashed) curves correspond to the contribution of the Borsellino ($\gamma-e^-$) diagrams. The asymmetries (84) are shown in the third row. The dotted curve in the upper left panel is the nondecreasing (with the increase of the photon-beam energy) contribution caused by the Borsellino diagrams that is defined by Eq.~(83). The upper indices in the second and third rows ($\pm$)
 denote the positive (+) or negative (-) value of the corresponding observables. For the negative value of the observable the logarithm of its modulus is shown.}\label{fig.4}}
\end{figure}

We see that the ($\gamma-e^-$) diagrams contribute  to the unpolarized part at 100~MeV$<\omega<$ 200~MeV
on the level of a few $\mu\,b$ (remind that we consider the $\varphi$-distribution, and integration over the angle $\varphi$ increases the unpolarized part $\sigma^{^U},$ as well as the polarization-dependent part $\sigma^{^{CL}},$ by a factor 2$\pi$). The Borsellino contribution to the total cross section is more than two orders larger.

%%%%%%%%%%%%%%%%%%%%%%%%%%%%%%%%%%%%%%%%%%%%%%%%

As regards the polarization-dependent parts at these energies, the ($\gamma-e^-$) contribution is the largest for the circularly polarized photon and the longitudinally polarized (along the direction ({\bf q})) target electron (CL-part). It is negative and amounts to  about
0.5~$\mu$b in the absolute value at $\omega=$~100~MeV, which very slowly decreases with the growth of the photon-beam energy. If the
the target electron is polarized transversally (in the plane $({\bf q\,,p_2})$) (CT-part), the corresponding part of the cross section is positive and equals to about 5$\cdot$10$^{-2}\mu\,b$ at $\omega=$~100~MeV and decreases up to 2$\cdot$10$^{-2}\mu\,b$ at $\omega=$~200~MeV. In the first case, the Borsellino contribution exceeds the ($\gamma-e^-$) one (in absolute value) by an order, and in the second case -- by two orders. The part, caused by the linear polarization of the photon (L-part), is negative for both the ($\gamma-e^-$) and Borsellino contributions (this agrees with previous calculations). The ($\gamma-e^-$) contribution is less than one percent of the Borsellino one at $\omega=$~100~MeV and decreases with the rise of the energy whereas the Borsellino contribution increases.

 Thus, to separate the effect due to the ($\gamma-e^-$) diagrams using the total cross section, even in the frame of the pure QED, the radiative corrections (RC) to the Borsellino contribution have to be taken into account. At present, we know such RC to the positron spectrum and to the total cross section in the Weizsacker-Williams approximation only \cite{VKM74} (in unpolarized case). This correction covers the region $q^2_m\leq m^2$ and amounts to about one percent. It means that for our goal we have to compute RC more accurately and consider in the first order the contribution of the region of the $q^2_m\sim$ a few tens of m$^2$ and even the leading (enhanced by a large logarithm) terms of the second order. Such calculations are absent at present.

%%%%%%%%%%%%%%%%%%%%%%%%%%%%%%%%%%%Gena%%%%%%%%%%%%%%%%%%%%%%%%%%%%%%%%%%%%%%%%%%%%%%%%%%%

%From Fig. 4 one can see that
The asymmetry $A^L$ (caused by the
linear polarization of the photon beam), as a function of the photon
energy $\omega $, is noticeable and depends weakly on $\omega $. The
contribution of the $\gamma -e^-$ diagrams is small ($\sim 1 \%$ at
$\omega \leq 10$ MeV) and decreases rapidly as the photon energy
increases. The asymmetry $A^{CL}$ is of the order of 10-15$\%$ in the region $\omega \leq
20$ MeV and decreases as a function of $\omega $. The contribution
of the $\gamma -e^-$ diagrams is less than 1$\%$ in the region
$\omega \leq 10$ MeV and decreases very rapidly. The behavior of the
asymmetry $A^{CT}$ is similar to the asymmetry $A^{CL}$ but it is
somewhat less. The circular polarization of the photon beam can be
determined, in principle, by means of the measurement of the
asymmetry $A^{CL}$ for the photon energies $\leq 20$ MeV.

%%%%%%%%%%%%%%%%%%%%%%%%%%%%%%%%%%%%%%%%%%%%%%%%%%%%%%%%%%%%%%%%%%%%%%

The main contribution to the triplet total cross section, given by Eq.~(83), arises due to the region of a very small values of  $q^2_m\sim m^2.$ To reduce this contribution kinematically, we have to select events with large values of $q^2_m\gg m^2.$
The different distributions over the $q^2_m$ variable (at large enough values of $q^2_m$) and the ratio R of the Borsellino contribution to the ($\gamma-e^-$) one are shown in Figs.~5,~6, and the polarization asymmetries defined by Eq.~(84) are given in Fig.~7.
\begin{figure}[t]
 \centering
\includegraphics[width=0.28\textwidth]{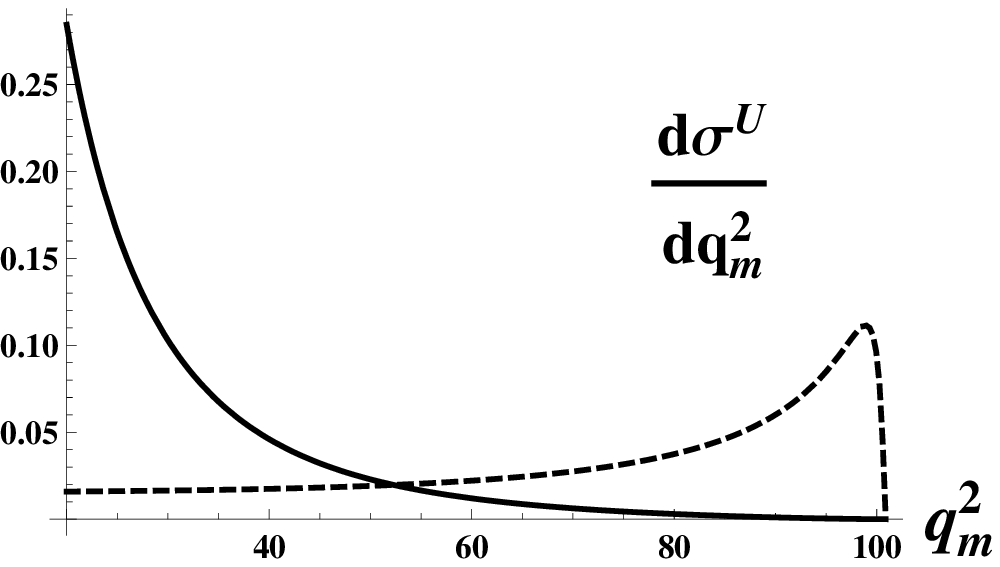}
\hspace{0.4cm}
\includegraphics[width=0.28\textwidth]{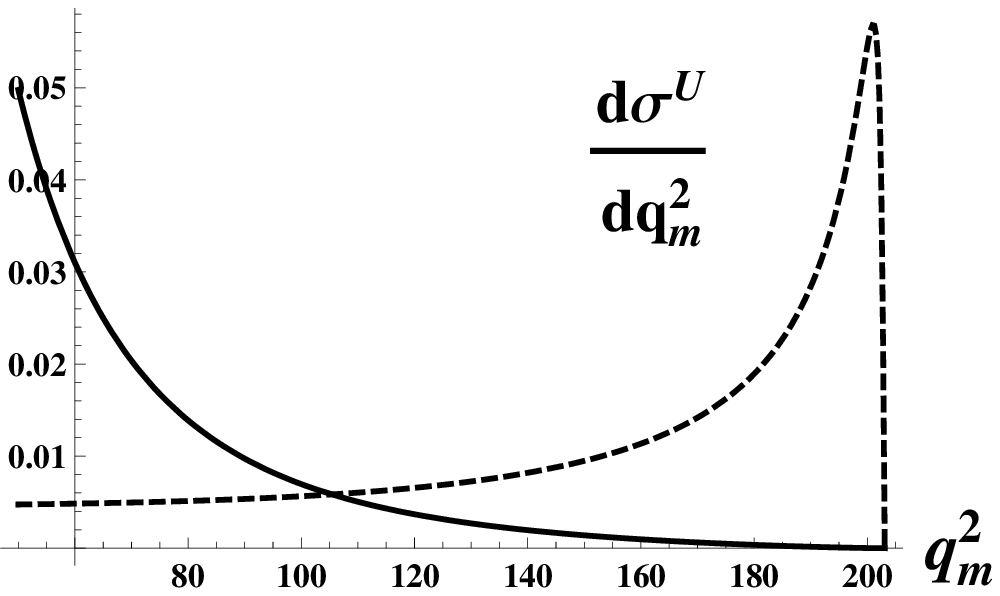}
\hspace{0.4cm}
\includegraphics[width=0.28\textwidth]{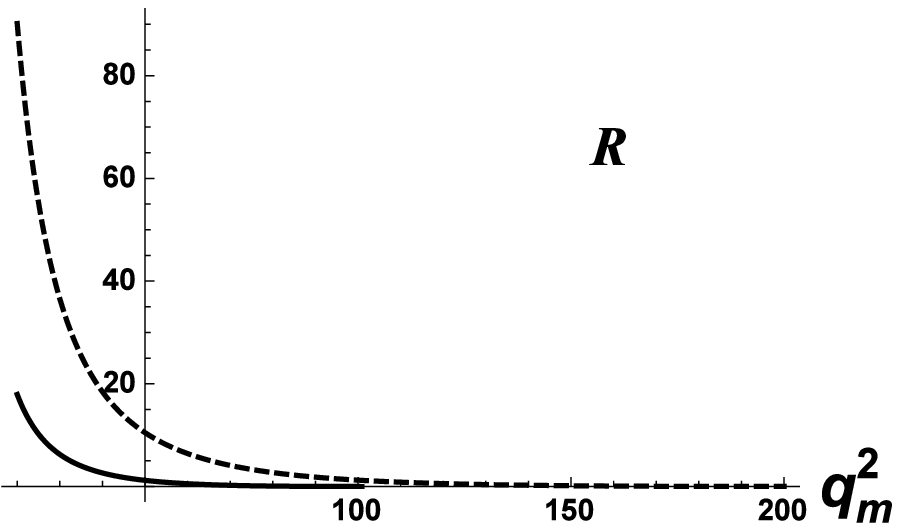}

\vspace{0.5cm}
\includegraphics[width=0.28\textwidth]{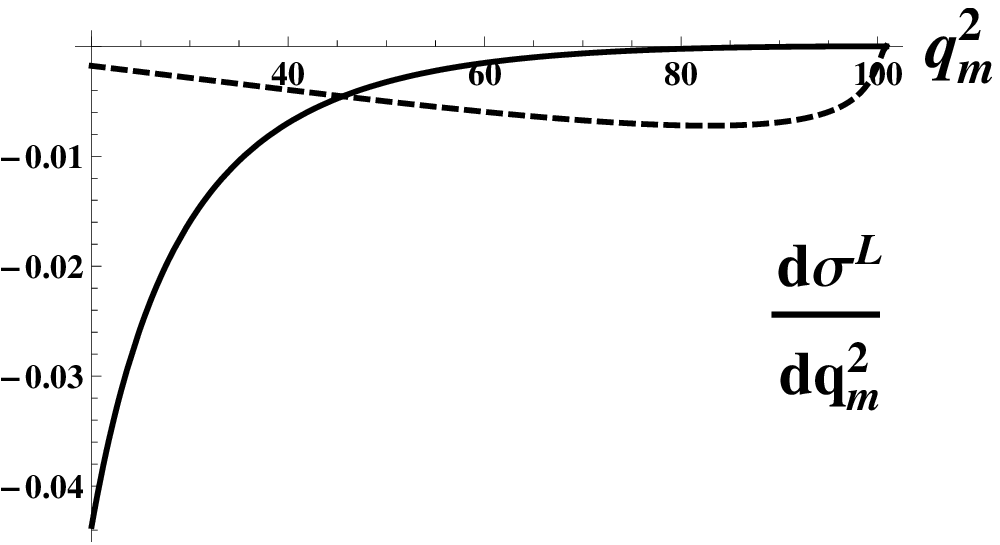}
\hspace{0.4cm}
\includegraphics[width=0.28\textwidth]{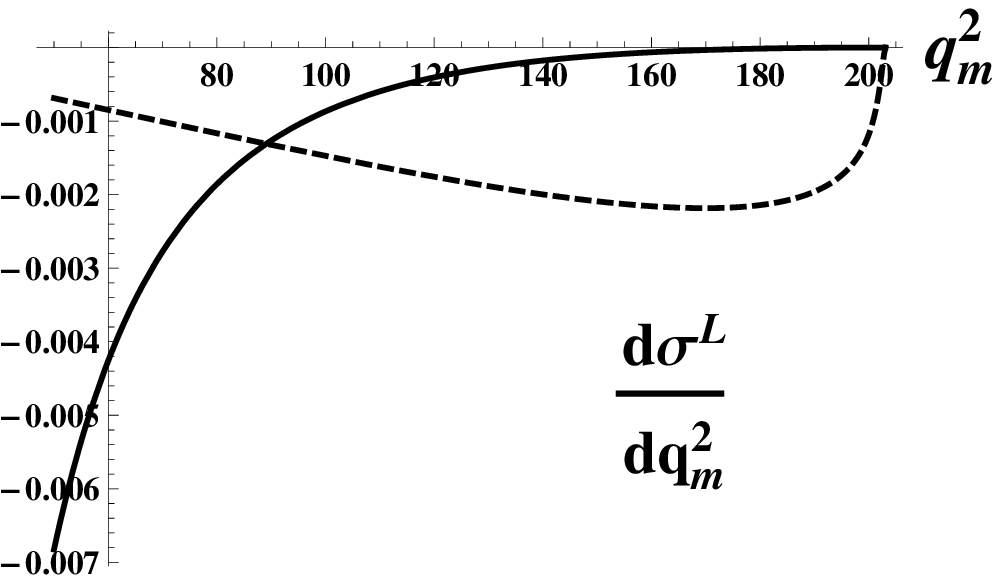}
\hspace{0.4cm}
\includegraphics[width=0.28\textwidth]{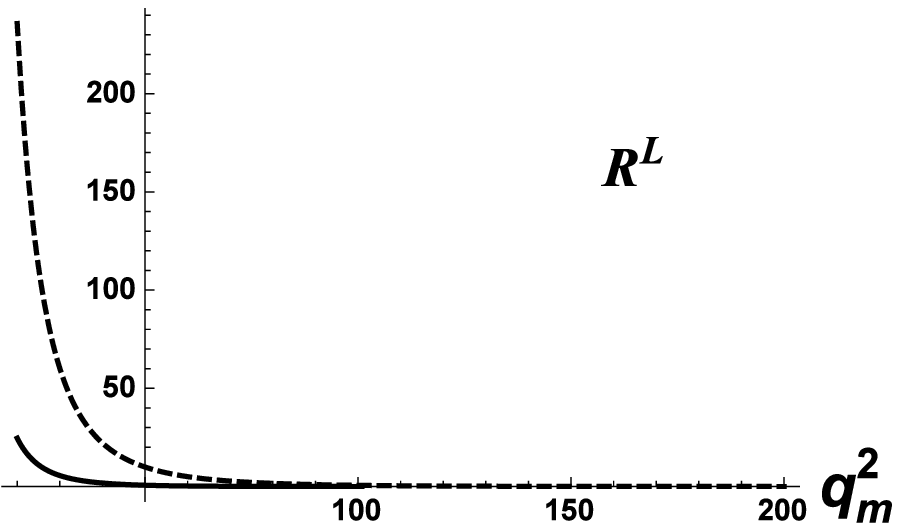}
\parbox[t]{1\textwidth}{\caption{The parts of the differential cross section (in $\mu\,b\cdot$MeV$^{-2}$) for the photon-beam energy 100~MeV and 200~MeV versus $q^2_m$ (in MeV$^2$) integrated over all the possible values of the created pair invariant mass squared $Q^2$. The upper row is the polarization independent part, and the lower one corresponds to the linearly polarized photon. The solid curves in the left and middle panels describe the Borsellino contribution and the dashed curves -- the ($\gamma-e^-$) one. In the right panels we show
ratio R and R$^L$ of the Borsellino contributions to the $(\gamma-e^-)$ one at $\omega=100$~MeV (solid curves) and 200~MeV (dashed curves).}\label{fig.5}}
 \end{figure}

\begin{figure}[t]
 \centering
\includegraphics[width=0.28\textwidth]{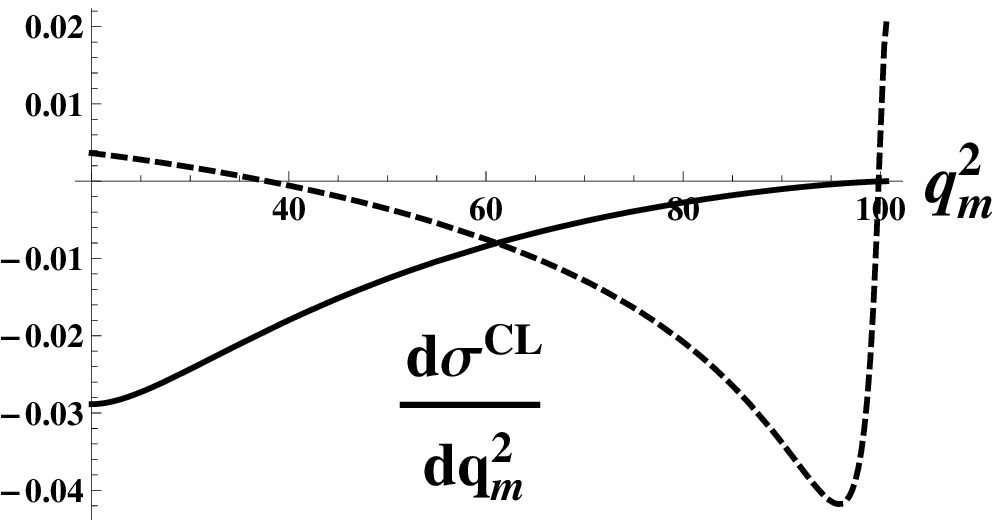}
\hspace{0.4cm}
\includegraphics[width=0.28\textwidth]{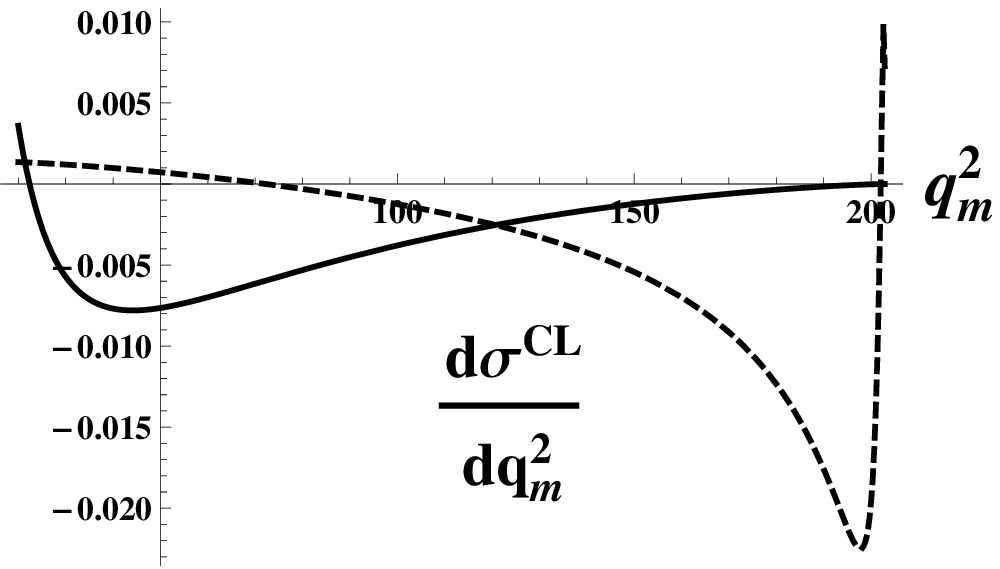}
\hspace{0.4cm}
\includegraphics[width=0.28\textwidth]{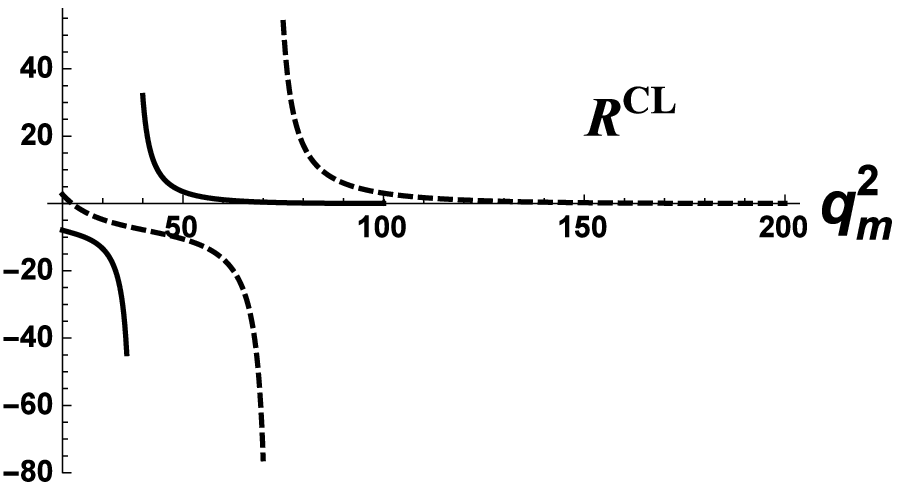}

\vspace{0.5cm}
\includegraphics[width=0.28\textwidth]{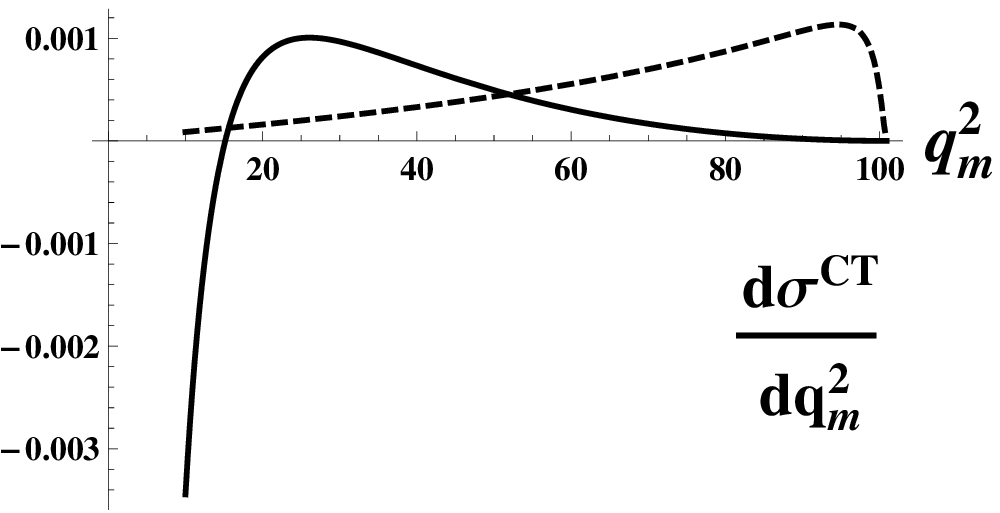}
\hspace{0.4cm}
\includegraphics[width=0.28\textwidth]{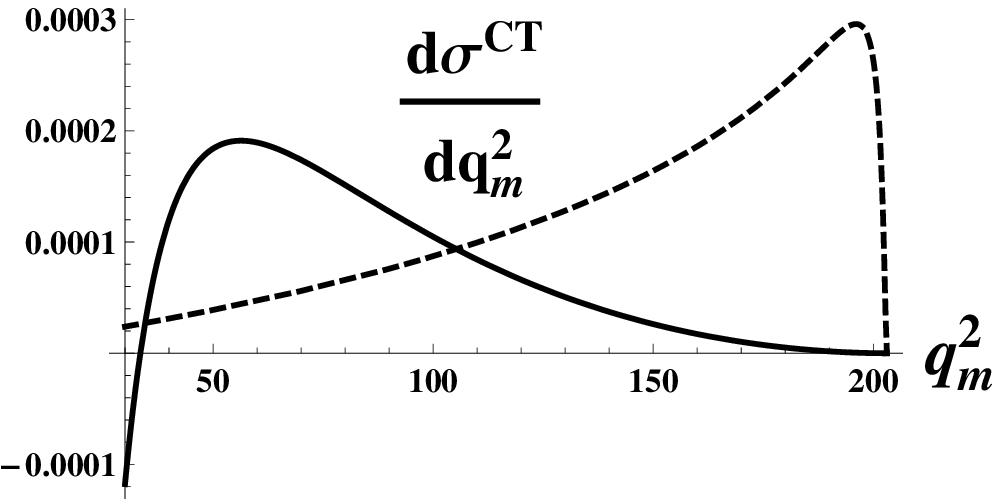}
\hspace{0.4cm}
\includegraphics[width=0.28\textwidth]{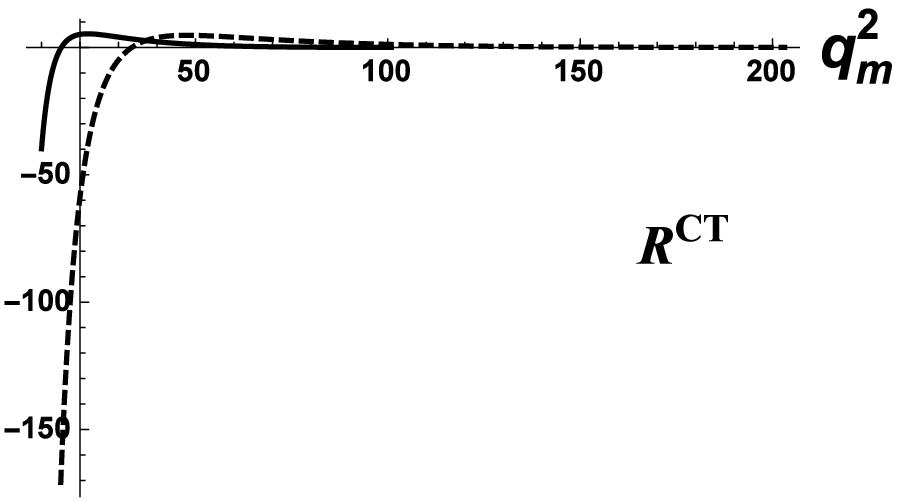}
\parbox[t]{1\textwidth}{\caption{The same as in Fig.~5 but for the case of the circularly polarized photon and longitudinally (the upper row) and transversely (the lower row) polarized electron.}\label{fig.6}}
 \end{figure}

\begin{figure}[t]
 \centering
\includegraphics[width=0.28\textwidth]{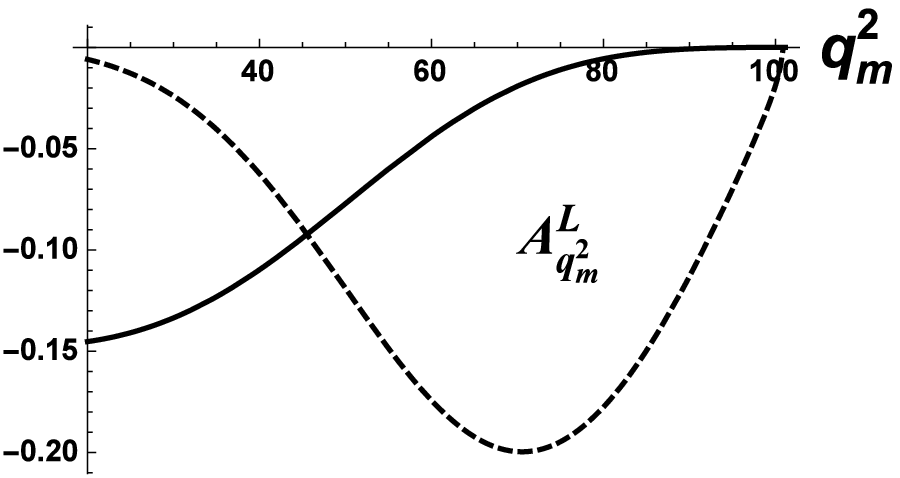}
\hspace{0.4cm}
\includegraphics[width=0.28\textwidth]{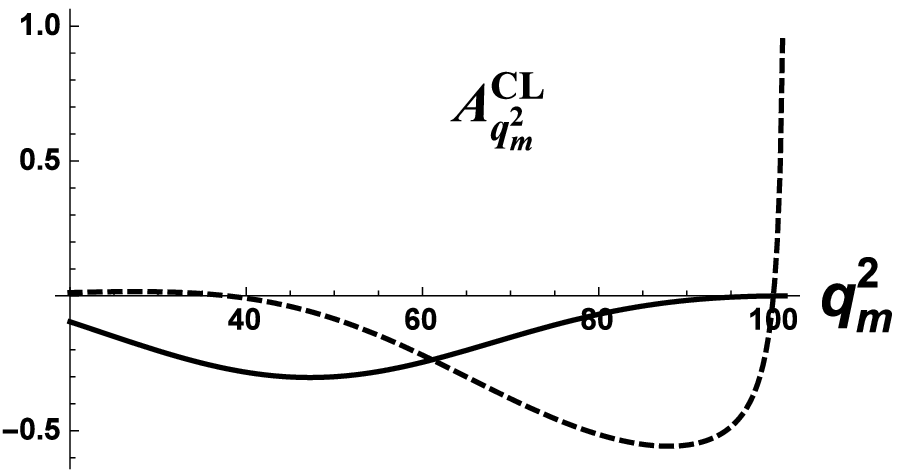}
\hspace{0.4cm}
\includegraphics[width=0.28\textwidth]{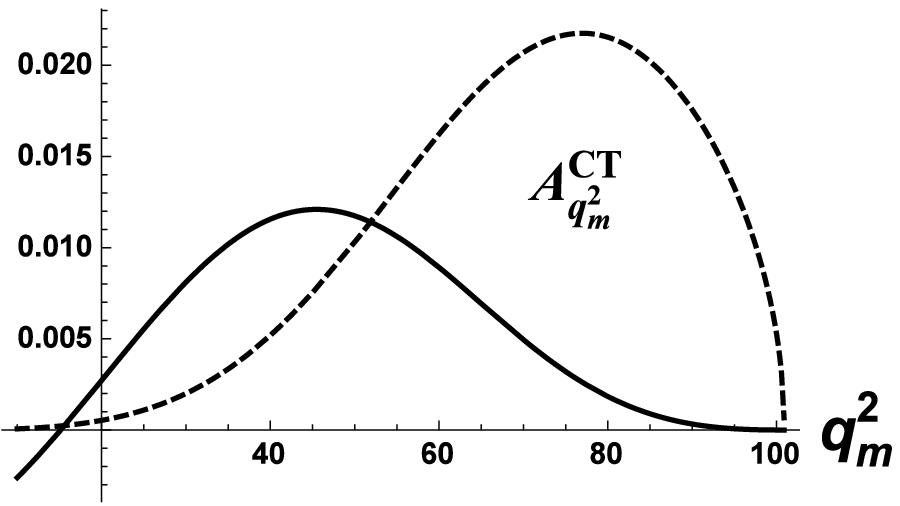}

\vspace{0.5cm}
\includegraphics[width=0.28\textwidth]{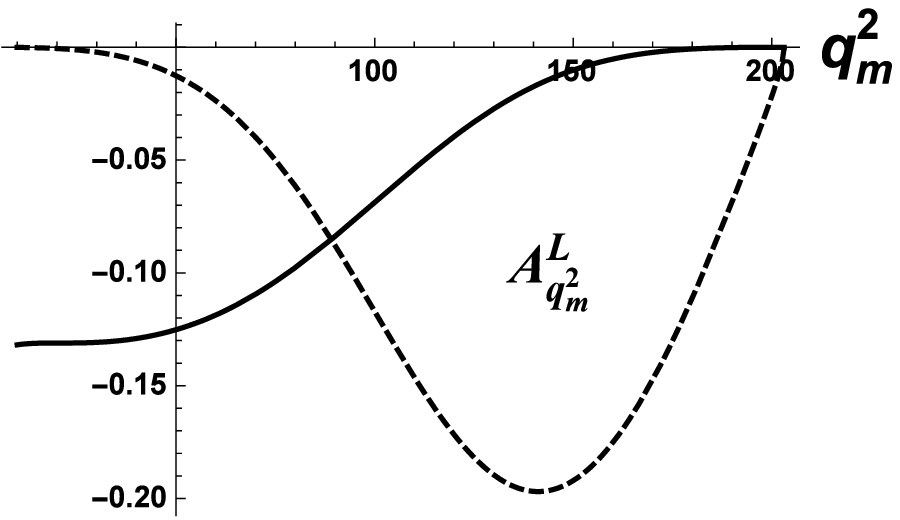}
\hspace{0.4cm}
\includegraphics[width=0.28\textwidth]{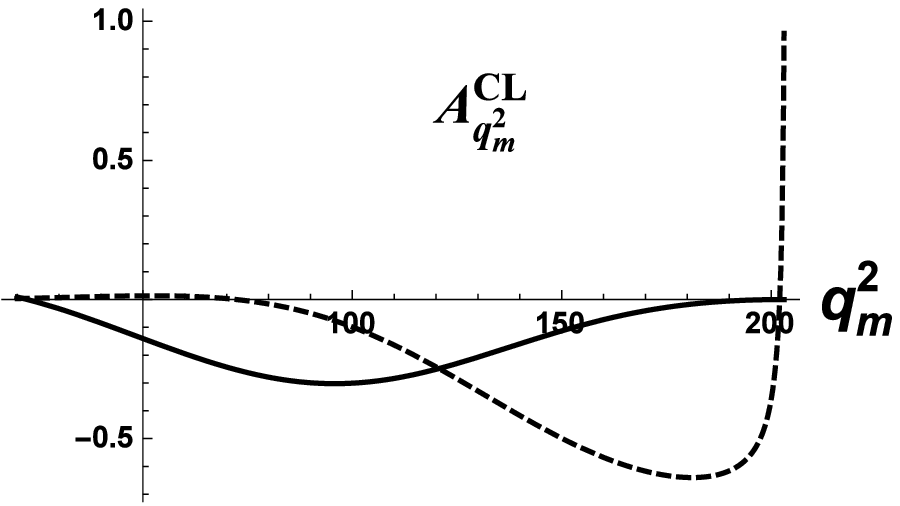}
\hspace{0.4cm}
\includegraphics[width=0.28\textwidth]{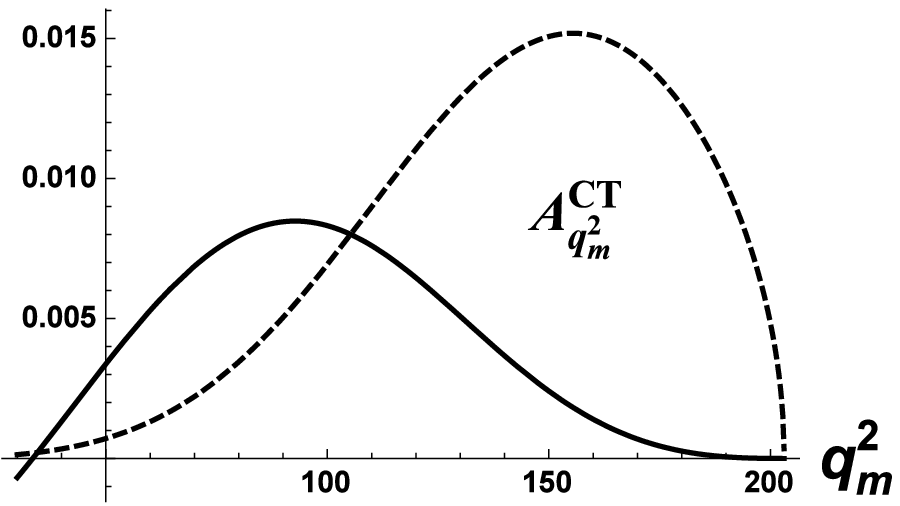}
\parbox[t]{1\textwidth}{\caption{The differential asymmetries as function of the $q^2_m$ variable at $\omega=$100~MeV (the upper row) and $\omega=$200~MeV (the lower row). The solid (dashed) curves describe the Borsellino ($(\gamma-e^-)$)
contribution. }\label{fig.7}}
\end{figure}

We see from Figs.~5,~6 that the ($\gamma-e^-$) contribution increases with the rise of $q^2_m$ whereas the Borsellino one decreases. Such behaviour ensures  the suppression of the Borsellino contribution at large $q^2_m$ near its maximal possible values, where the corresponding parts of the cross section is determined, almost completely, by the ($\gamma-e^-$) contribution. The unpolarized part equals more than 0.1~$\mu$b at $\omega=$100~MeV and two times smaller at $\omega=$200~MeV. At $\omega=$100~MeV the CL-part reaches,
in absolute value, 0.04~$\mu$b, the CT-part -- 0.001~$\mu$b and the L-part -- about 0.01~$\mu$b.

The Fig.~7 shows that the asymmetry $A^L_{q^2_m}$ is of the order of
15-20$\%$ in the region $q^2_m \leq 90 (160)$ MeV$^2$ at $\omega
$=100 (200) MeV. The Borsellino contribution decreases and the
$\gamma -e^-$ one increases. The asymmetry $A^{CL}_{q^2_m}$ is of the
order of 40-50$\%$. The $\gamma -e^-$ contribution is small in the
region $q^2_m \leq 50 (100)$ MeV$^2$ at $\omega $=100 (200) MeV and
beyond this region its contribution is dominated. The magnitudes of
the asymmetry $A^{CT}_{q^2_m}$ is less than 1$\%$ in entire considered
region. We see that the asymmetry $A^{CL}_{q^2_m}$ is appreciable and can
be measured, in principle. So, it is possible to determine the
circular polarization of the photon beam at more higher energies
than for the case of the asymmetry $A^{CL}$ since the general
picture of the behavior of these asymmetries depends weakly on the
photon energy.

%%%%%%%%%%%%%%%%%%%%%%%%%%%%%%%%%%%%%%%%%%%%%%%%%%%%

To search for the deviation from SM due to the possible mixing of a photon with the light dark matter candidate (U-boson), it is necessary to study the distribution over the created pair invariant mass $Q^2$. In Figs.~8~-~11 we show such distributions which derived by the integration over the restricted region of $q^2_m$, including the events with $q^2_m>~$10, 20 and 40~MeV$^2$ and analyse the corresponding effects. In contrast to the Borsellino contribution, the $(\gamma-e^-)$ one depends not very strong on the value of the $q^2_m$ cut.

In Fig.~8 we show the unpolarized part of the cross section caused by the $(\gamma-e^-)$ diagrams with chosen restrictions and the ratio R$^{-1}$ of the $(\gamma-e^-)$ to the Borsellino contributions. We see that for $\omega$=100~MeV, even for the events with $q^2_m >$ 20~MeV$^2,$ there exist regions of the relative small and large values of $Q^2$ where the  $(\gamma-e^-)$ contribution is greater than the Borsellino one (but the event number is larger at small $Q^2$). For the events with $q^2_m >$ 40~MeV$^2,$ this effect manifests itself more  significantly.
The more sizeable restriction on the minimal values of $q^2_m$ is required at larger photon-beam energies to decrease the
Borsellino contribution, as it is seen even for $\omega$=200~MeV.

\begin{figure}[hbt]
 \centering
\includegraphics[width=0.35\textwidth]{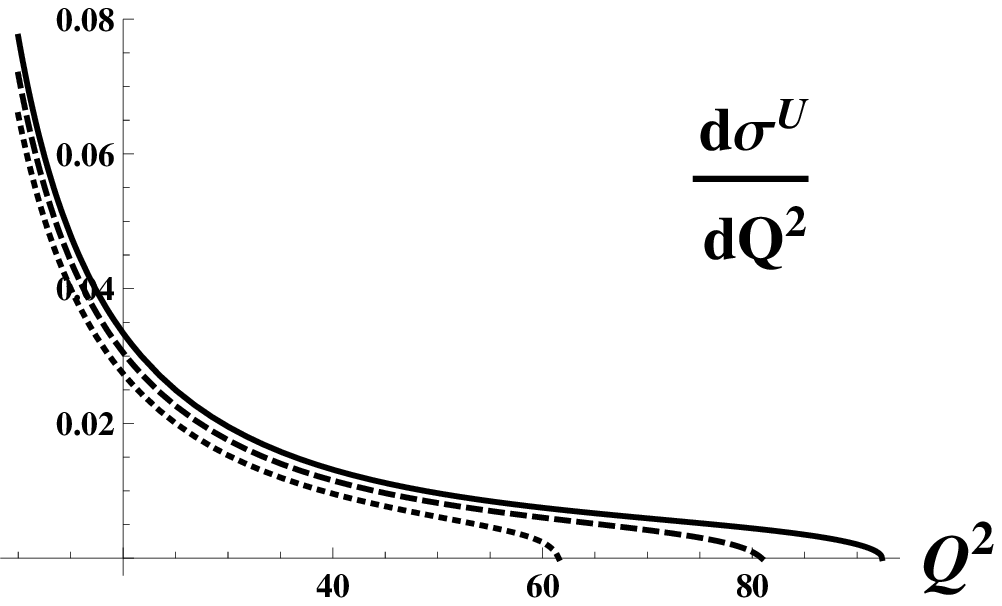}
\hspace{0.4cm}
\includegraphics[width=0.35\textwidth]{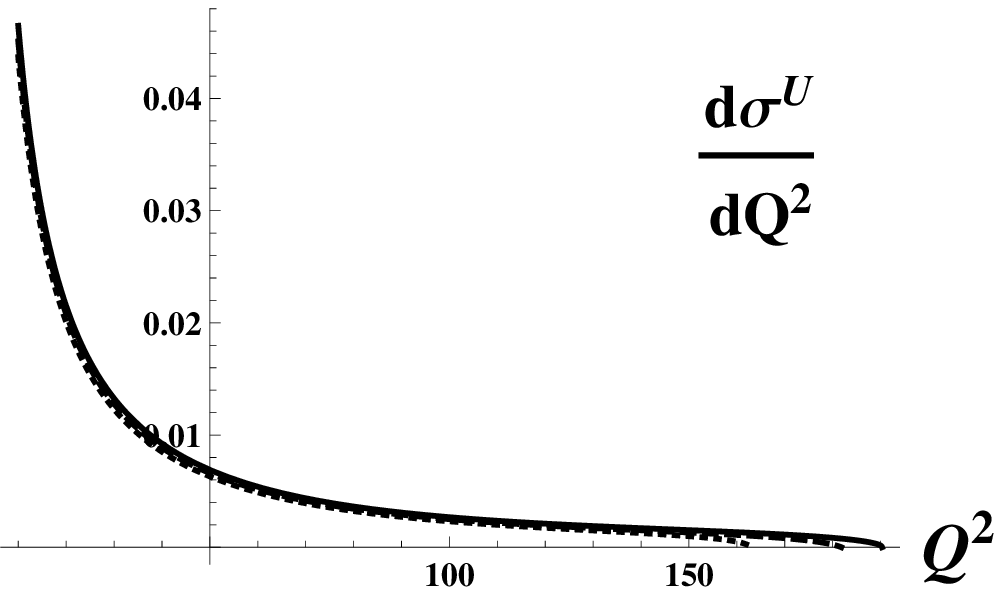}

\vspace{0.5cm}
\includegraphics[width=0.28\textwidth]{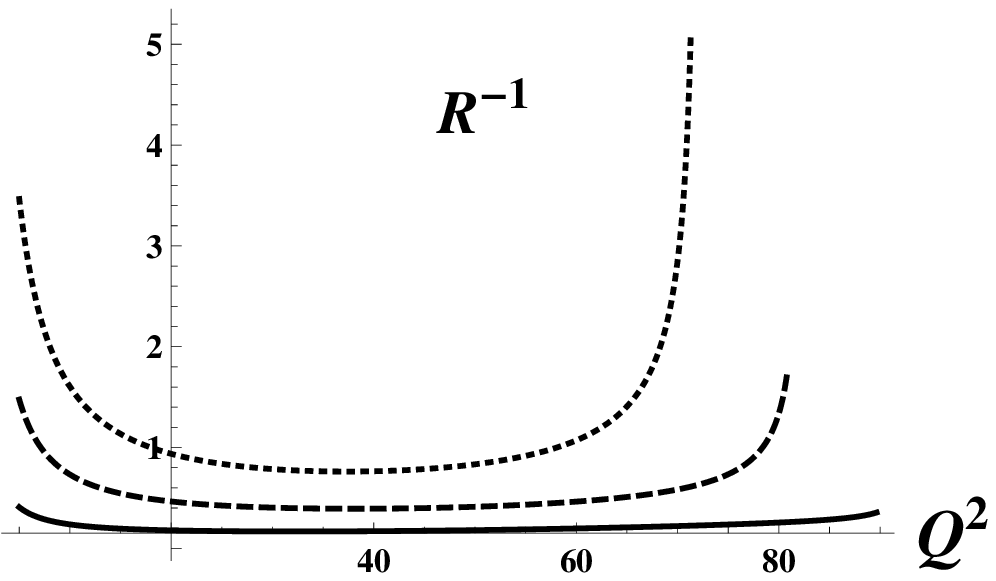}
\hspace{0.4cm}
\includegraphics[width=0.28\textwidth]{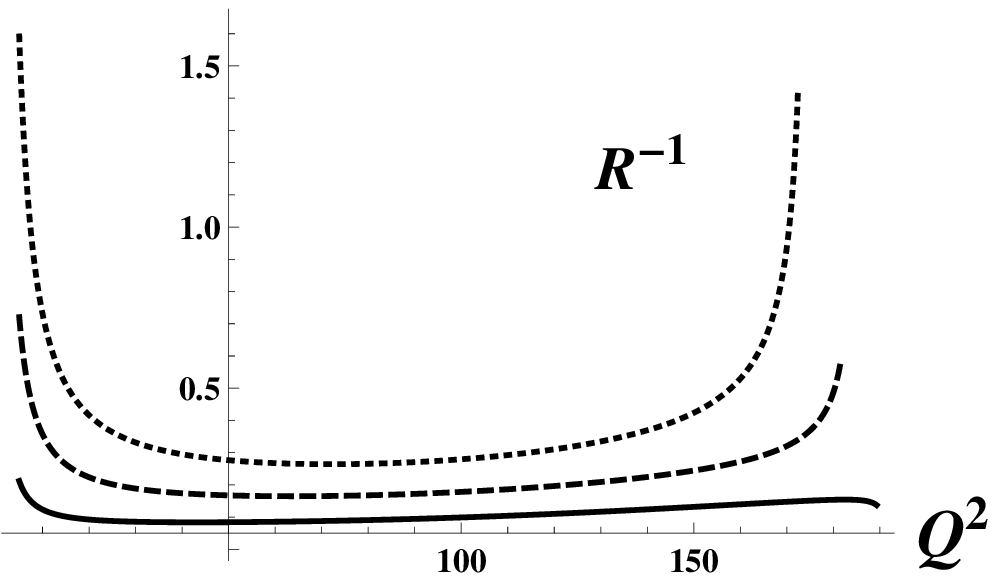}
\parbox[t]{1\textwidth}{\caption{The $(\gamma-e^-)$ contribution to the polarization-independent part of the cross section as a function of $Q^2,$ in $\mu\,b\cdot$MeV$^{-2},$ at different restrictions on the event selection (the upper row) at $\omega$=100~MeV (the left panel) and 200~MeV (the right panel). The solid line corresponds to the events with $q^2_m >$ 10~MeV$^2,$ the dashed line --  $q^2_m >$ 20~MeV$^2$ and the dotted one -- $q^2_m >$ 40~MeV$^2.$ In the lower row we show the respective ratio $R^{-1}$ of the $(\gamma-e^-)$ contribution to the Borsellino one.}\label{fig.8}}
 \end{figure}

In Figs.~9,~10 we show the effect caused by the restriction on the minimal values of $q^2_m$ for the polarized parts of the cross section for both the  $(\gamma-e^-)$ and Borsellino contributions.
The corresponding polarization asymmetries, defined by Eq.~(85), are given in Fig.~11. Again, we see that the $(\gamma-e^-)$
diagrams give the dominant contribution in the region of the large values of $Q^2$
and at large enough cuts on the $q^2_m$ variable.
\begin{figure}[hbt]
 \centering
\includegraphics[width=0.25\textwidth]{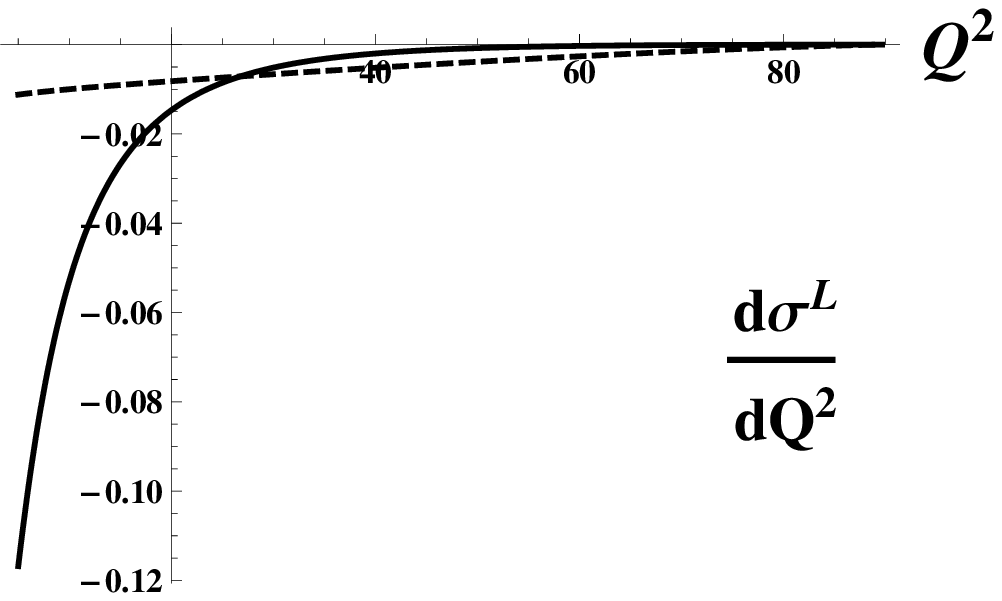}
\hspace{0.4cm}
\includegraphics[width=0.25\textwidth]{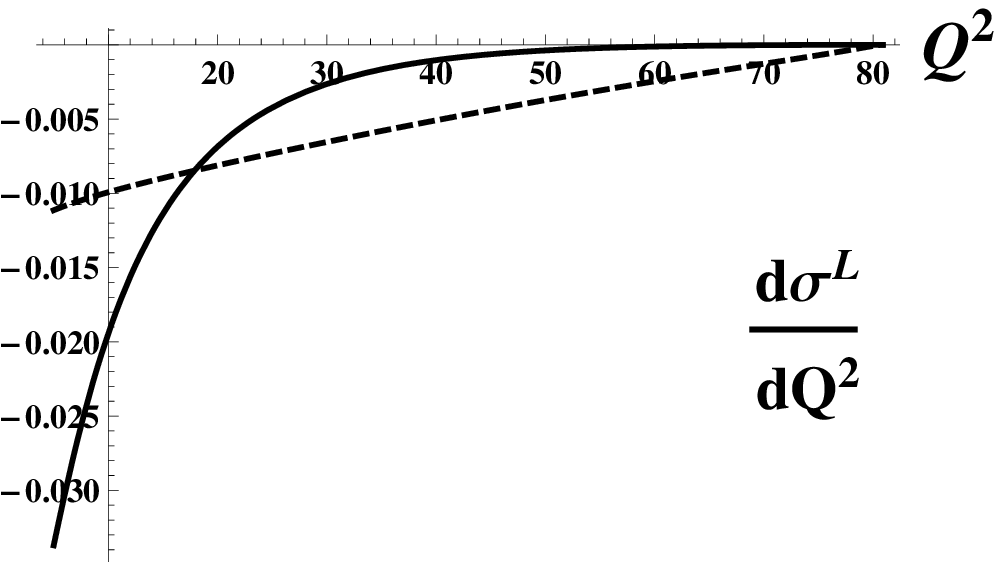}
\hspace{0.4cm}
\includegraphics[width=0.25\textwidth]{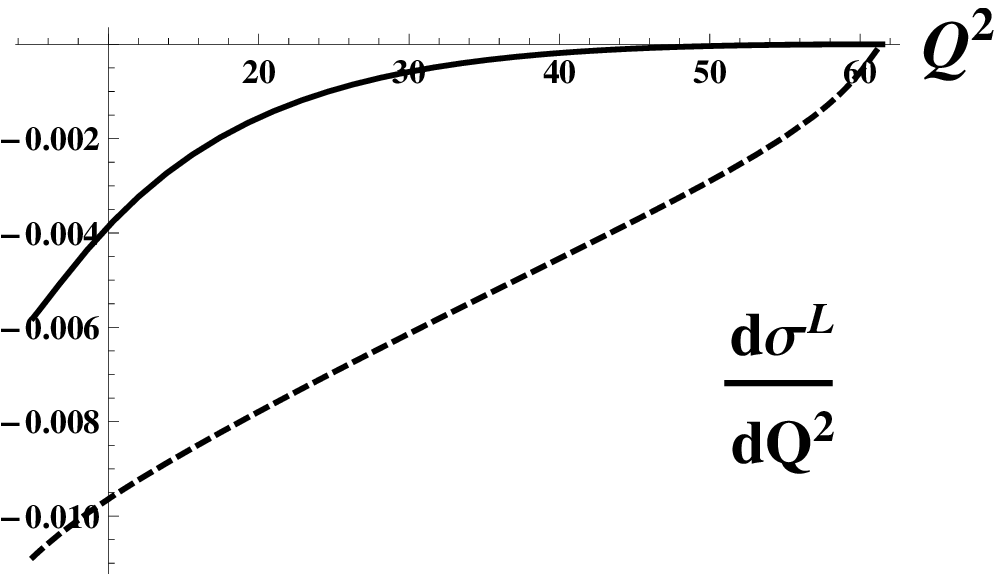}

\vspace{0.5cm}
\includegraphics[width=0.25\textwidth]{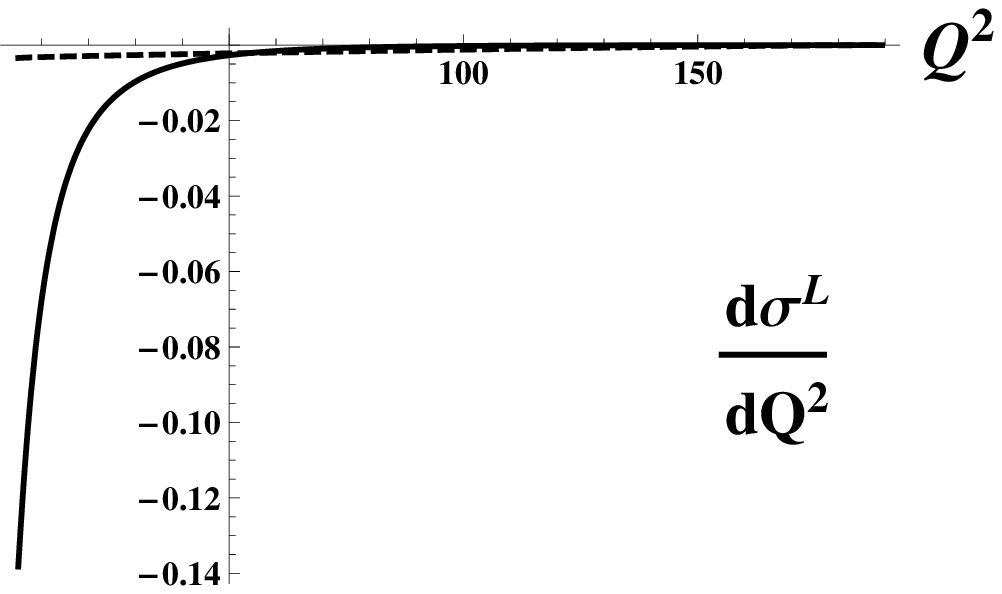}
\hspace{0.4cm}
\includegraphics[width=0.25\textwidth]{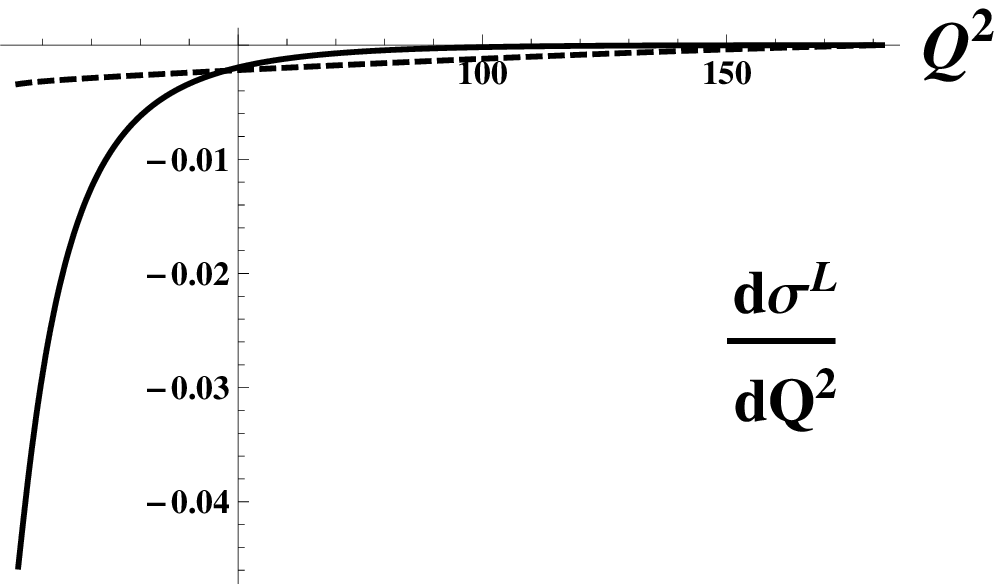}
\hspace{0.4cm}
\includegraphics[width=0.25\textwidth]{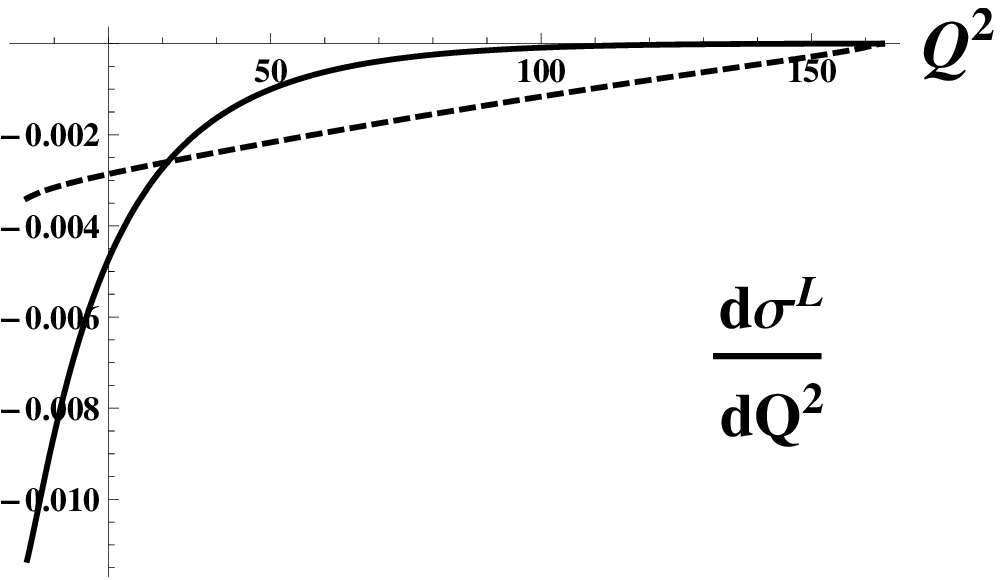}
\parbox[t]{1\textwidth}{\caption{The part of cross section which depends on the linear polarization of the photon, in $\mu\,b\cdot$MeV$^{-2},$ for $\omega$=100~MeV (the upper row) and $\omega=$ 200~MeV (the lower row). The left (middle, right) panel corresponds to the events with $q^2_m>$ 10 (20, 40) MeV$^2.$ The solid (dashed) line describes the Borsellino  ($(\gamma-e^-)$) contribution.}\label{fig.9}}
 \end{figure}

\begin{figure}[hbt!]
 \centering
\includegraphics[width=0.25\textwidth]{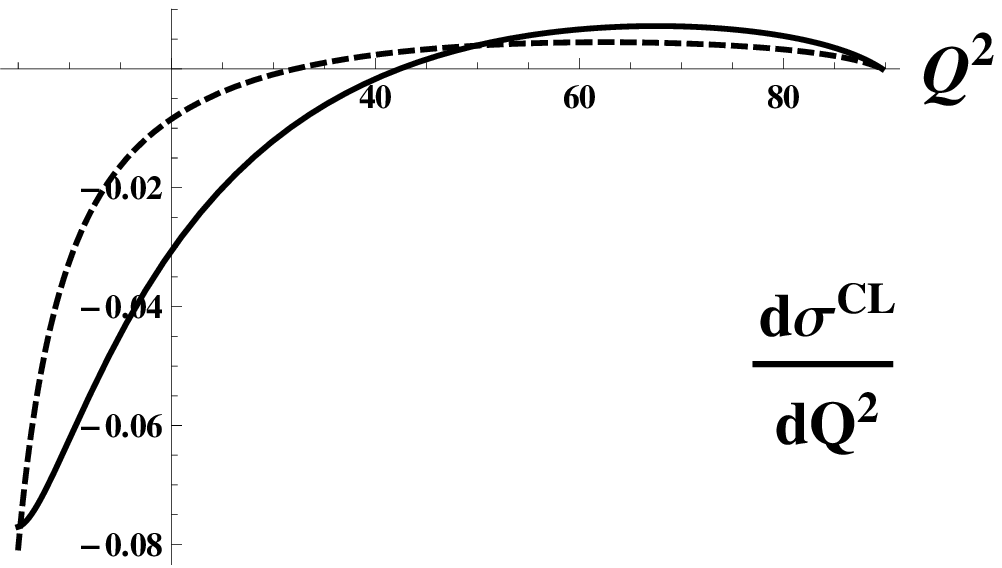}
\hspace{0.4cm}
\includegraphics[width=0.25\textwidth]{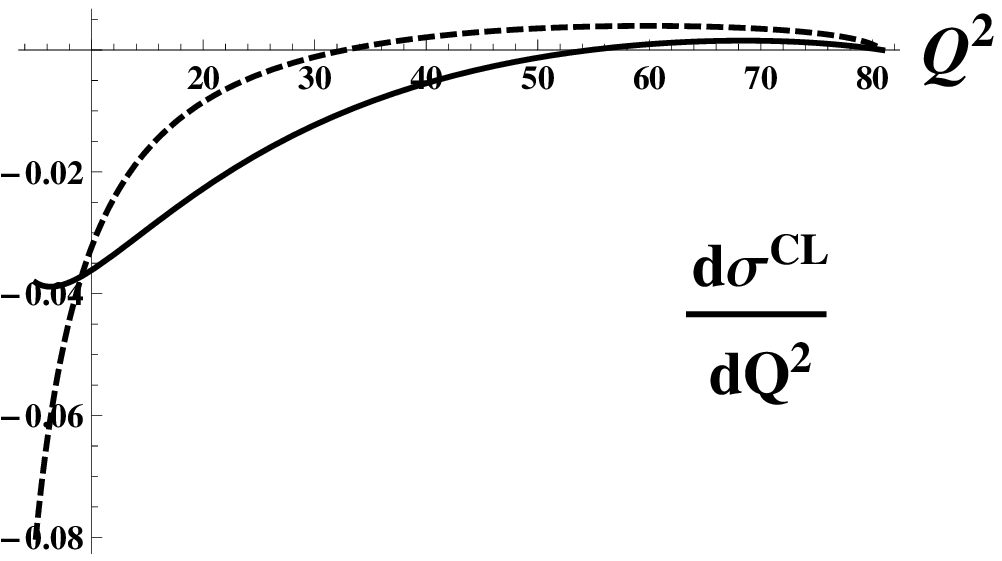}
\hspace{0.4cm}
\includegraphics[width=0.25\textwidth]{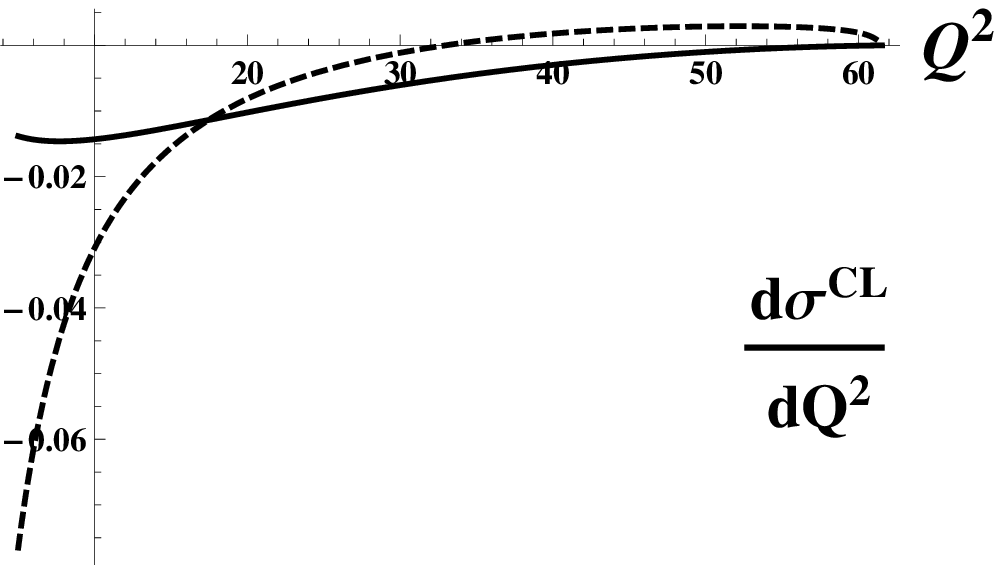}

\vspace{0.5cm}
\includegraphics[width=0.25\textwidth]{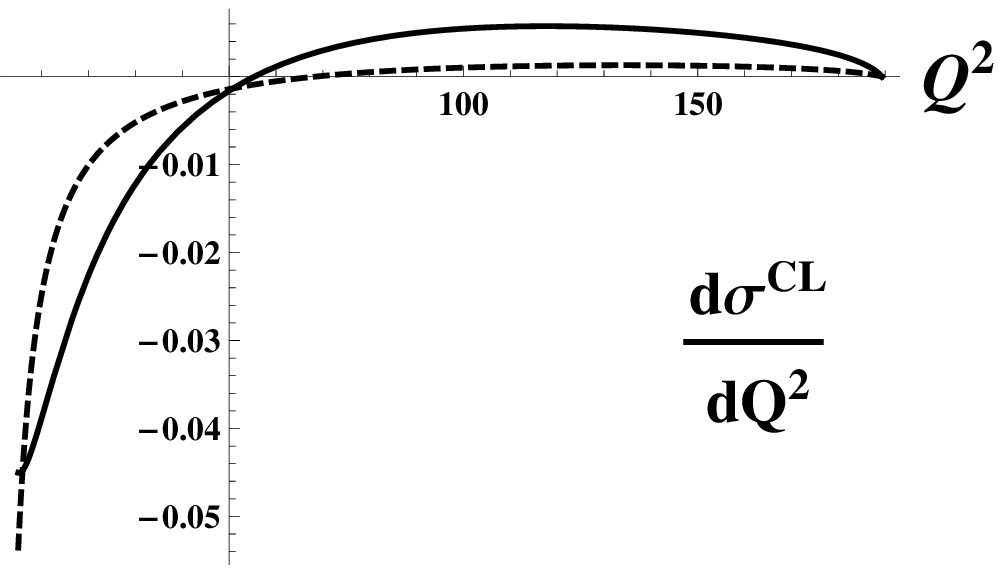}
\hspace{0.4cm}
\includegraphics[width=0.25\textwidth]{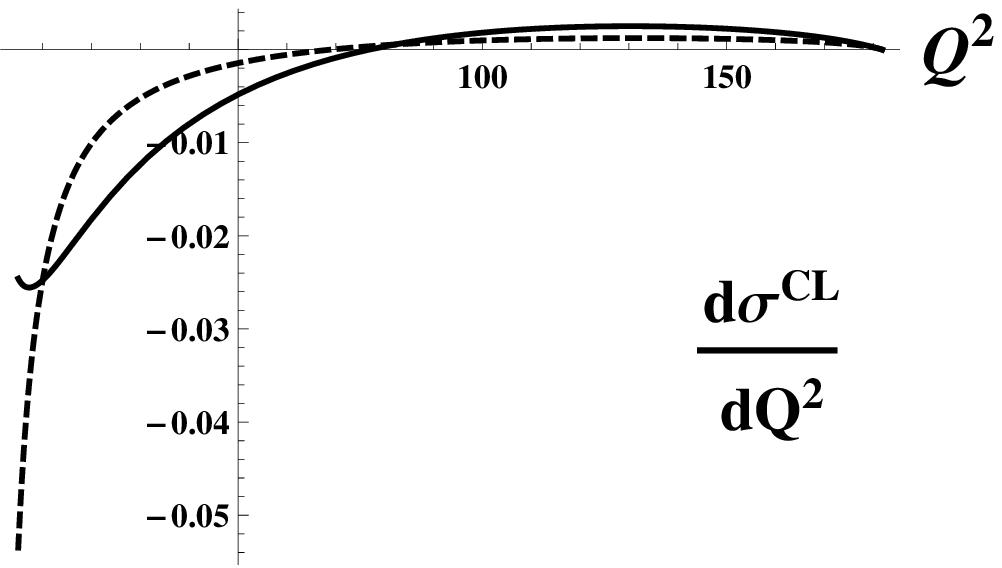}
\hspace{0.4cm}
\includegraphics[width=0.25\textwidth]{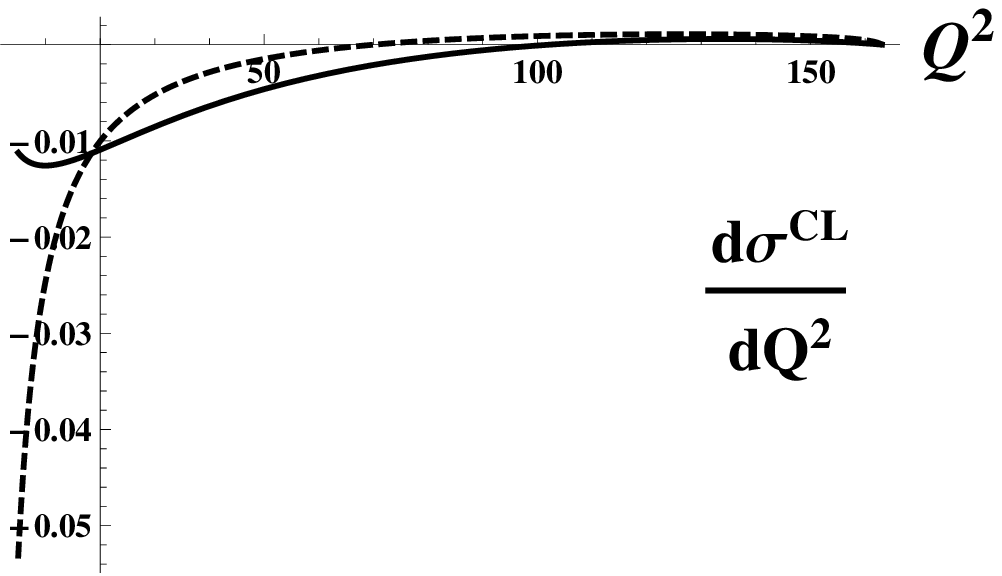}

\vspace{0.5cm}
\includegraphics[width=0.25\textwidth]{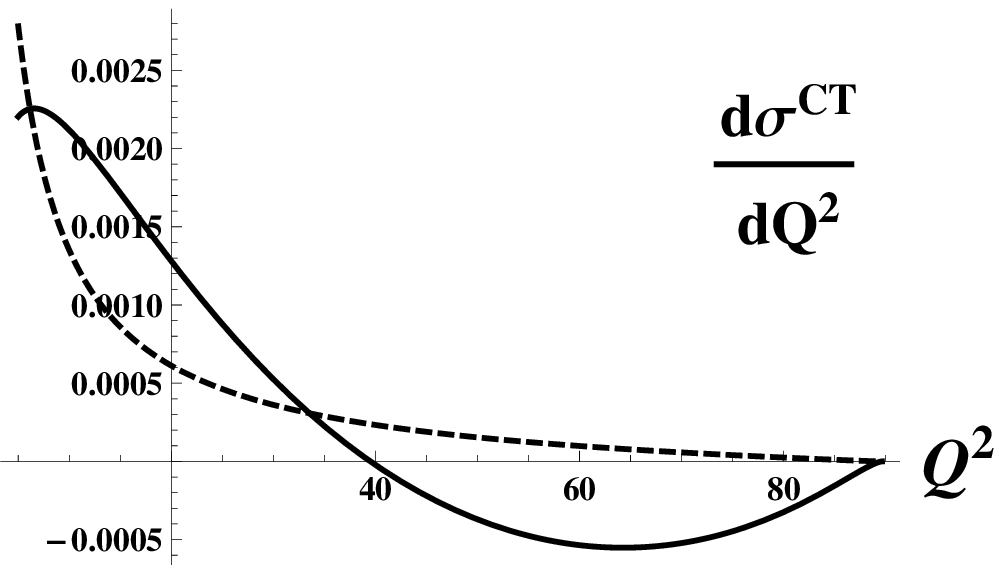}
\hspace{0.4cm}
\includegraphics[width=0.25\textwidth]{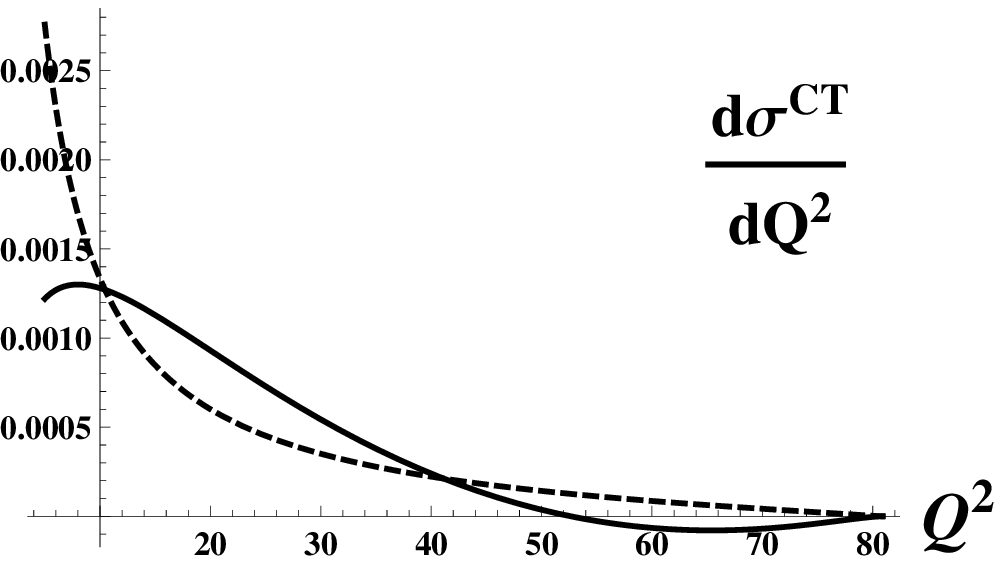}
\hspace{0.4cm}
\includegraphics[width=0.25\textwidth]{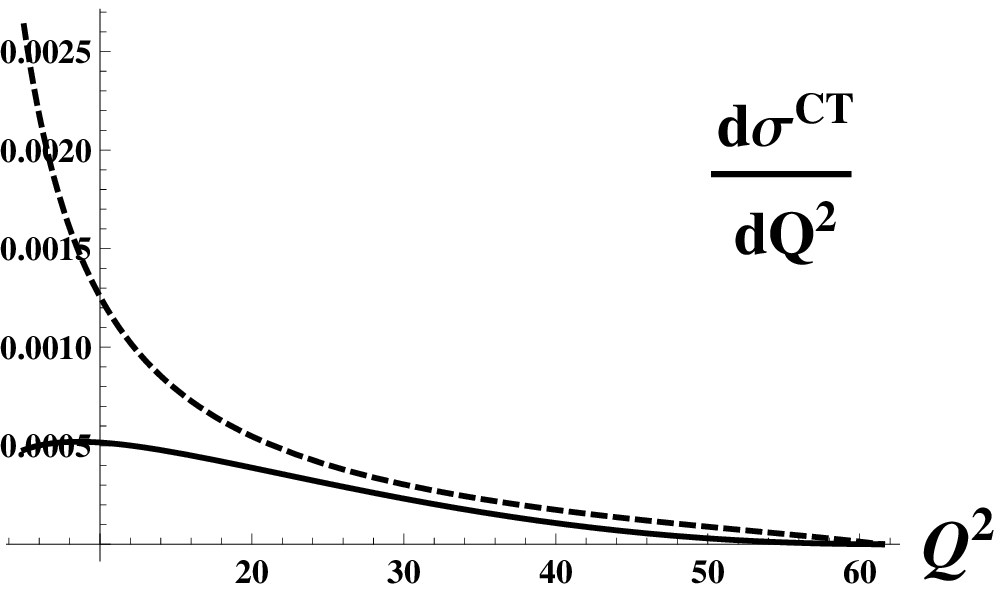}

\vspace{0.5cm}
\includegraphics[width=0.25\textwidth]{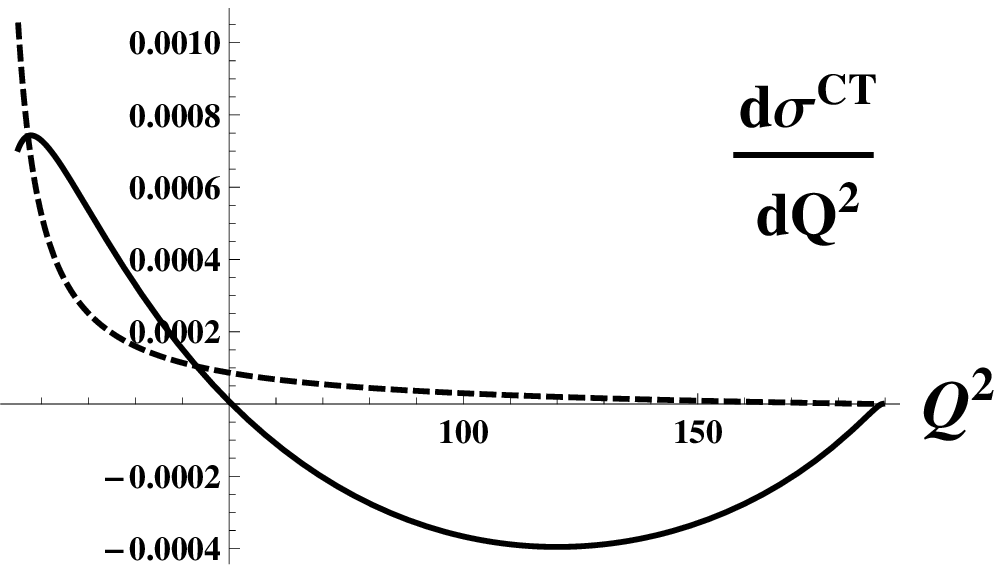}
\hspace{0.4cm}
\includegraphics[width=0.25\textwidth]{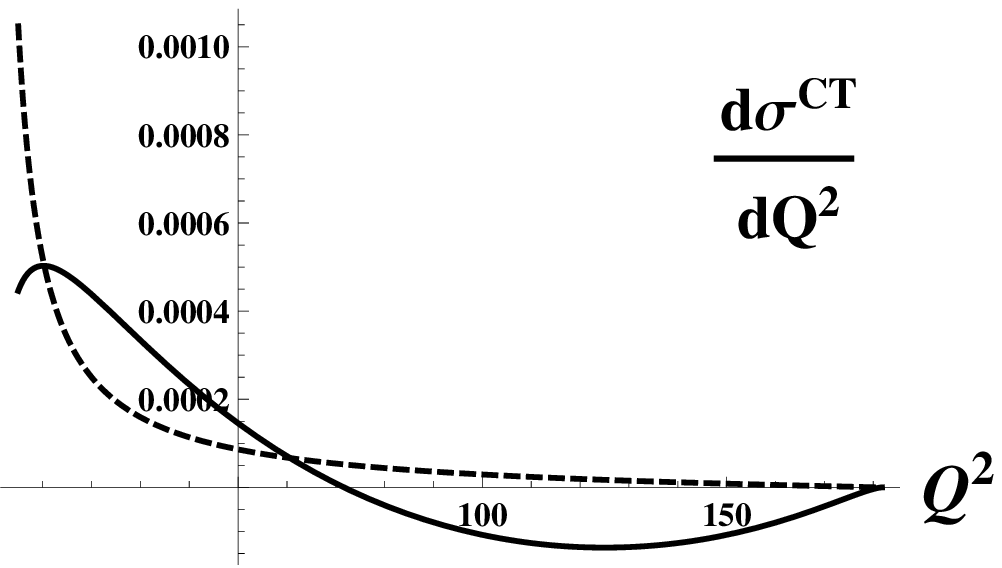}
\hspace{0.4cm}
\includegraphics[width=0.25\textwidth]{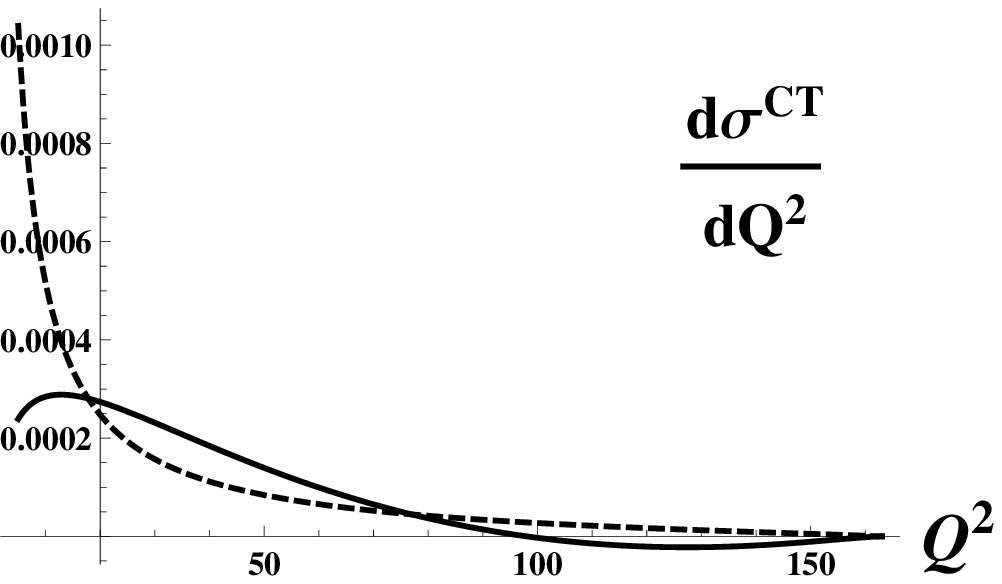}

\parbox[t]{1\textwidth}{\caption{The same as in Fig.~9 but for the parts of the cross section which depend on the circular polarization of the photon. The first and second rows correspond to the longitudinal polarization of the target electron, the third and fourth rows -- to the transversal one.}\label{fig.10}}
\end{figure}

\begin{figure}[hbt]
 \centering
\includegraphics[width=0.25\textwidth]{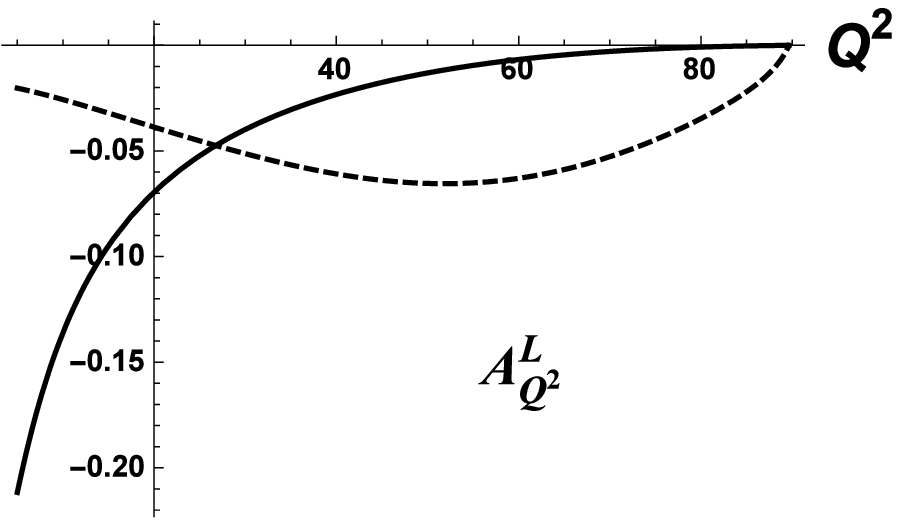}
\hspace{0.4cm}
\includegraphics[width=0.25\textwidth]{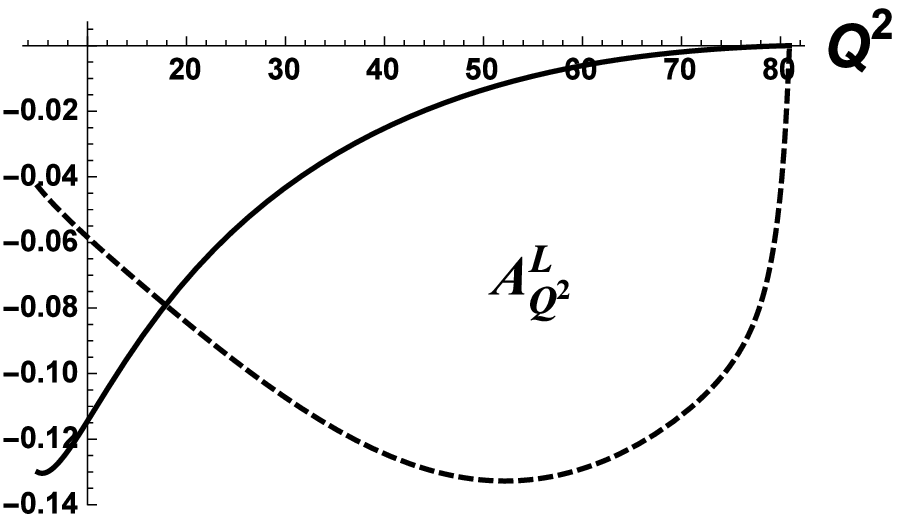}
\hspace{0.4cm}
\includegraphics[width=0.25\textwidth]{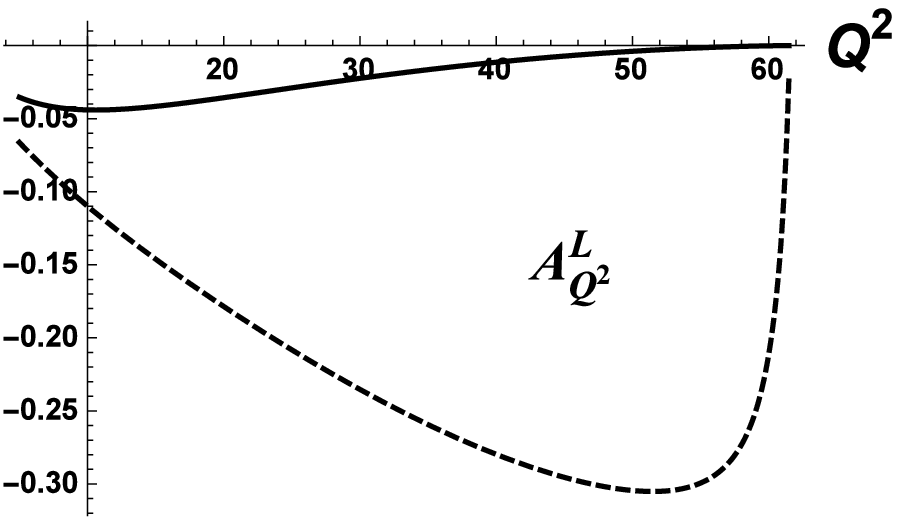}

\vspace{0.5cm}
\includegraphics[width=0.25\textwidth]{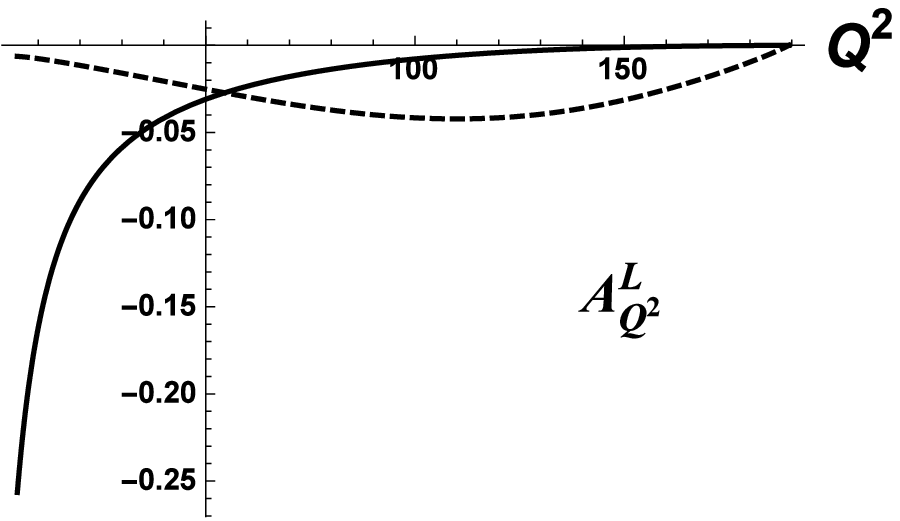}
\hspace{0.4cm}
\includegraphics[width=0.25\textwidth]{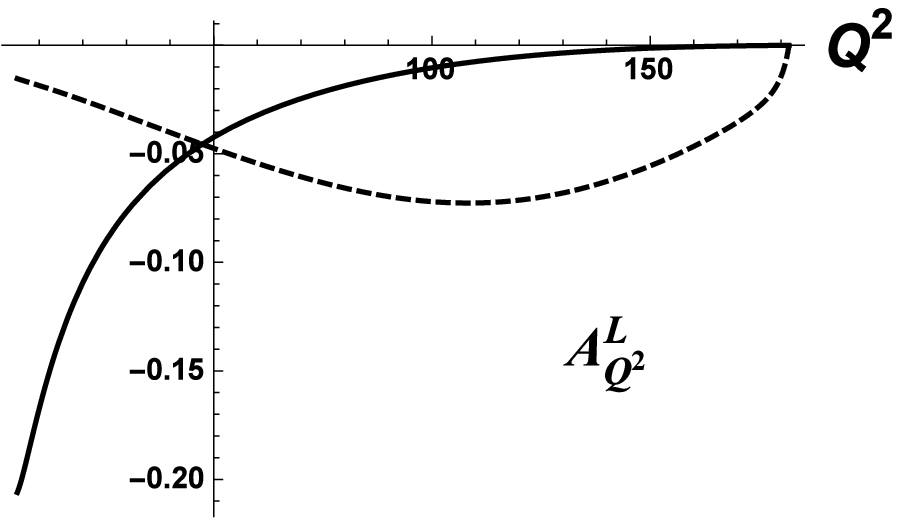}
\hspace{0.4cm}
\includegraphics[width=0.25\textwidth]{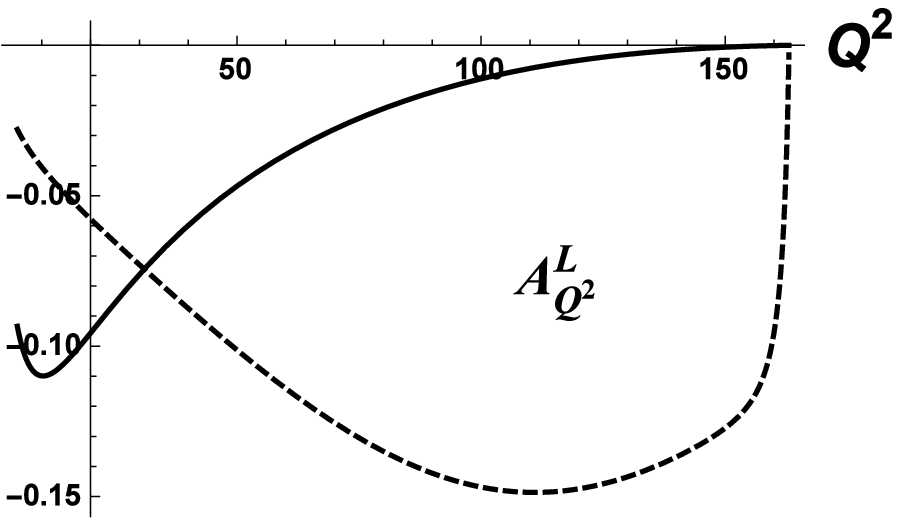}

\vspace{0.5cm}
\includegraphics[width=0.25\textwidth]{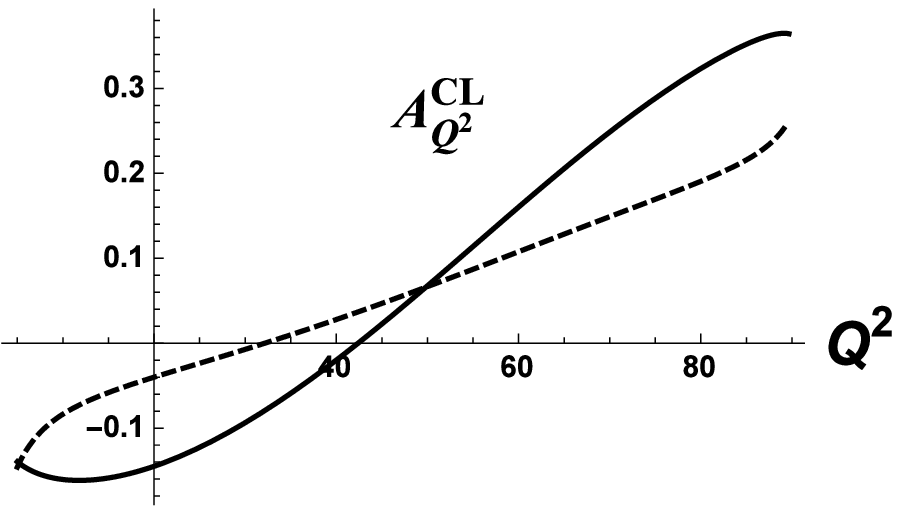}
\hspace{0.4cm}
\includegraphics[width=0.25\textwidth]{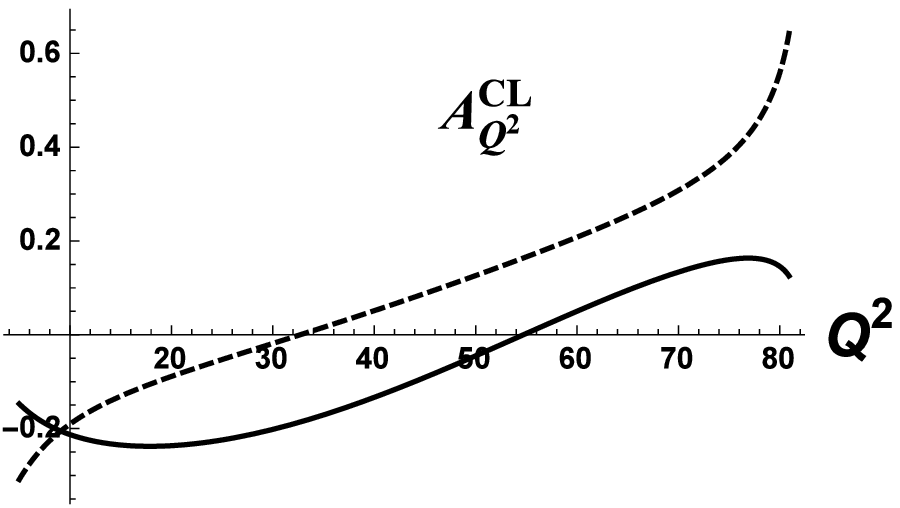}
\hspace{0.4cm}
\includegraphics[width=0.25\textwidth]{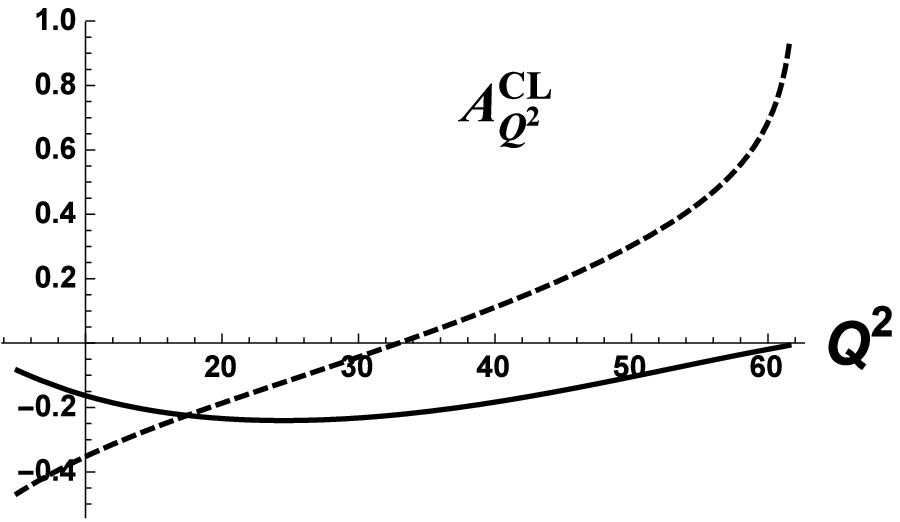}

\vspace{0.5cm}
\includegraphics[width=0.25\textwidth]{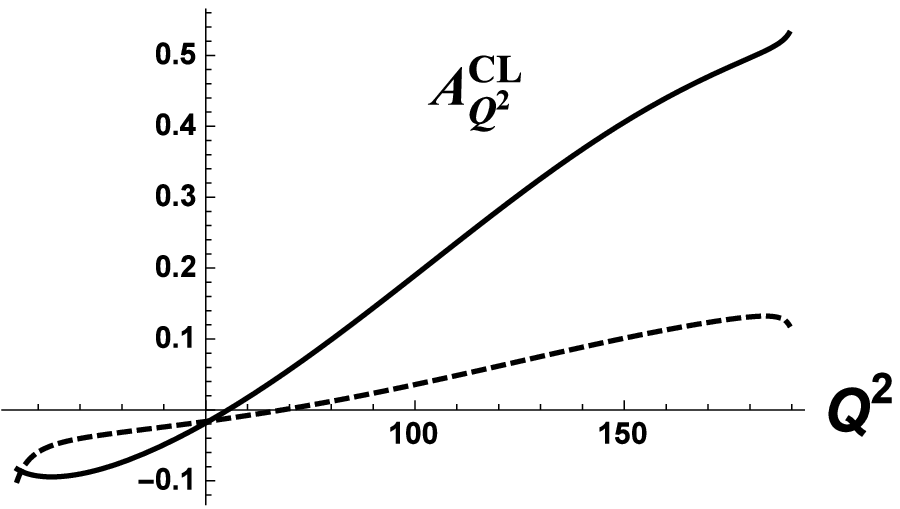}
\hspace{0.4cm}
\includegraphics[width=0.25\textwidth]{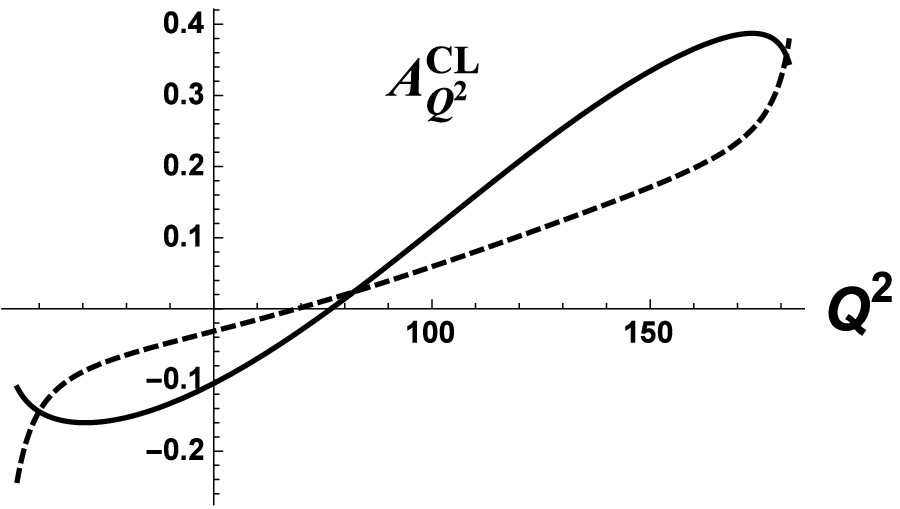}
\hspace{0.4cm}
\includegraphics[width=0.25\textwidth]{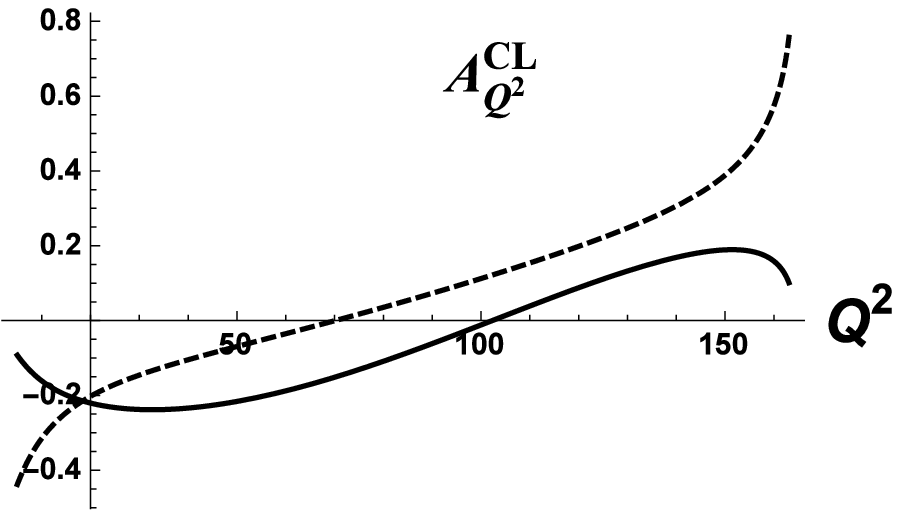}

\parbox[t]{1\textwidth}{\caption{The L$-$  and CL$-$asymmetries as a functions of the created $e^+e^-$-pair invariant mass squared $Q^2$ at $\omega=$100~MeV (the first and third rows) and $\omega=$200~MeV (the second and fourth rows).  The left (middle, right) panels correspond to the events with $q^2_m<$10 (20, 40)~MeV$^2$. The dashed (solid) lines describe the ($\gamma-e^-$) (Borsellino) contribution.}\label{fig.11}}
\end{figure}

\section{Conclusion}

In this paper, we analyzed the process of the triplet
photoproduction on a polarized electron target by a polarized photon
beam. The calculation of various observables has been done in the
approach when four Feynman diagrams were taken into account. Besides
the two Borsellino diagrams we took into account the two ($\gamma
-e^-$) (or Compton-like) diagrams. So, we neglect the effects of the
final electron identity. The results obtained in a such
approximation describe the events with well separated created and
recoil electrons. Otherwise, it is necessary to take into account
the identity effects. The numerical calculations were performed in
the laboratory system for the photon energy less than 200~MeV.

We investigated the possibility of determining the circular
polarization degree of a high energy photon by measuring the
asymmetry in the triplet production by a circularly polarized photon
beam on a polarized electron target. The influence of the
($\gamma-e^-$) diagrams contribution on the calculated observables are also analysed.

We think that the measuring the distribution over the invariant mass
of the created electron-positron pair in the process of the triplet
photoproduction would be a good method to search for a dark photon.
So, we search for the kinematical regions  where the contribution of
the dominated background mechanism (the Borsellino diagrams) can be
suppressed as compared with the useful Compton-like diagrams where
the signal from the dark photon may be measured. Besides, the
contribution of the Borsellino diagrams can be calculated with
the necessary accuracy.

For the first time, the different distributions were obtained in the
analytical form. We obtain the double distribution over the $q^2$
(the square of the four-momentum transfer to the recoil electron)
and $Q^2$ (the created $e^+e^-$-pair invariant mass squared)
variables, and single distributions over the $q^2$ or $Q^2$
variables.

%%%%%%%%%%%%%%%%%%%%%%Gena%%%%%%%%%%%%%%%%%%%%%%%%
We obtain the expressions for the asymmetry $A^L$ caused by the
linear polarization of the photon beam as a function of the photon
energy $\omega $ and as a function of $q^2_m$ at fixed $\omega $.
The asymmetries $A^{CL}~(A^{CT})$, caused by the circularly polarized
photon beam and polarized initial electron in the case when the
polarization vector of the target is parallel (orthogonal) to the
photon momentum, have been also calculated as a functions of $q^2_m$
and $\omega $. Although the final expressions for all the
observables are given in the laboratory system but until this step all
the expressions are given in the invariant form and it is easy to
transform the final expressions to the invariant form by the
substitution $\omega =(W^2-2m^2)/2m$.

It was found that the measurement of the $A^{CL}$ or $A^{CL}_{q^2_m}$
asymmetries can be used for the determination of the circular
polarization of the photon beam for the $\omega \leq 20~(200)$ MeV
for the case of $A^{CL}~(A^{CL}_{q^2_m})$.

%%%%%%%%%%%%%%%%%%%%%%%%%

\section{Acknowledgments}
%%%%%%%%%%%%%%%%%%%%%%%%%%
This work was partially supported by the Ministry of Education and
Science of Ukraine (project no. 0115U000474).

\end{document}